\documentclass[aps,prl,reprint,superscriptaddress
]{revtex4-2}
\usepackage[labelformat=simple]{subcaption}
\usepackage{graphicx}
\usepackage{dcolumn}
\usepackage{bm}

\usepackage{amssymb}
\usepackage{amsmath}
\usepackage{hyperref}
\usepackage{physics}
\usepackage{booktabs}
\usepackage{esint}
\usepackage{soul}
\usepackage{cancel}
\usepackage{kotex}
\usepackage[margin=1in]{geometry}
\usepackage{array}
\usepackage{amssymb}  
\usepackage{siunitx}
\begin{document}

\preprint{APS/123-QED} 

\title{Dissipative Channels Determine Open Electromagnetic Quantization}

\author{Hyunwoo Choi}
\thanks{These authors contributed equally to this work.}
\affiliation{Department of Electrical Engineering, Pohang University of Science and Technology (POSTECH), Pohang 37673, Republic of Korea}

\author{Junwoo Gim}
\thanks{These authors contributed equally to this work.}
\affiliation{Department of Electrical Engineering, Pohang University of Science and Technology (POSTECH), Pohang 37673, Republic of Korea}

\author{Thomas E. Roth}
\affiliation{Elmore Family School of Electrical and Computer Engineering, Purdue University, West Lafayette, Indiana 47907, USA}

\author{Weng Cho Chew}
\affiliation{Elmore Family School of Electrical and Computer Engineering, Purdue University, West Lafayette, Indiana 47907, USA}

\author{Dong-Yeop Na}
\email{dyna22@postech.ac.kr}
\affiliation{Department of Electrical Engineering, Pohang University of Science and Technology (POSTECH), Pohang 37673, Republic of Korea}
\affiliation{Elmore Family School of Electrical and Computer Engineering, Purdue University, West Lafayette, Indiana 47907, USA}

\begin{abstract}
We formulate a quantization scheme for open electromagnetic systems with arbitrary passive boundary conditions.
Rather than specifying reservoirs phenomenologically, the method identifies them from the dissipation geometry of the Maxwell operator. 
Factoring the imaginary part of the Maxwell operator gives a bosonic realization of the field operator and separates the fluctuation channels into medium-assisted reservoirs from material absorption and boundary-assisted reservoirs from exchange through the open boundary. 
Depending on the boundary condition, the latter become free-space radiation modes, impedance-load channels, guided port modes, or more general boundary channels. 
Green-function input-output relations then follow as an application, yielding frequency-dependent scattering and noise kernels without Markov or single-mode assumptions. 
To illustrate the practical application, we consider a lossy structure with mixed impedance and outgoing boundaries, and photonic integrated circuit configurations with waveguide port boundaries.
\end{abstract}

\maketitle
\section{Introduction}
Contemporary quantum electrodynamic (QED) systems are inherently open architectures~\cite{breuer_OQS,PRX_master,wiseman2010quantum}. 
Photonic integrated circuits~\cite{PIC_nature,Jahani_NatComm_2018,Spencer_Optica_2014}, plasmonic structures~\cite{Plasmonics_PRL_2014,Plasmonics_nature}, waveguide devices~\cite{Sheremet_RMP_waveguideQED_2023,Melati_JOpt_2014}, microwave networks~\cite{Microwave_PRL,Microwave_nature,circuit_QED_RMP}, cavity QED nodes~\cite{cavityQED_RMP,cavityQED_nature}, and quantum transducers~\cite{transducer_IOP,transducer_nature} exchange energy with external degrees of freedom (DoFs) through radiation leakage, impedance loads, guided ports, and circuit terminations, even in the absence of appreciable material absorption.
At the quantum level, these channels are not merely classical loss mechanisms; they define the reservoirs required to preserve field commutation relations and determine the fluctuation spectrum of the open system.
Macroscopic QED (MQED)~\cite{Scheel2008MacroscopicQED,Dung2000QEDLocalized,Gruner1996QEDEvanescent} accounts for absorptive media by expressing the field operator in terms of local Langevin noise currents associated with material loss. 
A material-noise representation alone, however, is not a complete channel decomposition of an open quantum system.

The modified Langevin noise formalism (MLNF)~\cite{ciatonni_MLN_2024,Na2023quantumEMLossy,ciattoni_MLN_2026} resolved this issue by introducing boundary-assisted fluctuation associated with outgoing plane-wave scattering modes~\cite{Choi_PRApplied,Choi_FDTDQE,Choi_twoquanta,Forestire_PRA_2025,Forestire_nanophotonics_2025}. 
This construction is well suited to radiating nanoparticles and bulk free-space scattering problems~\cite{ciattoni_scattering_v1,ciattoni_scattering_v2,Weisha_HOM}, where the exterior reservoir is the homogeneous radiation continuum. 
However, in many structured electromagnetic environments, the external DoFs are not plane waves at infinity, but channels selected by engineered boundaries. 
These may be guided modes of a waveguide port~\cite{Moon2024TAP}, dissipative modes of an impedance load, coaxial or circuit ports, exact Dirichlet-to-Neumann (DtN) maps, or more general passive boundary operators~\cite{TAP_BC}. 
Such systems require a quantization principle in which the reservoir channels are determined by the imposed Maxwell boundary condition, rather than by a prescribed free-space radiation basis.

Here, we show that the quantum channels for quantization are determined by the dissipative geometry of the Maxwell operator. 
For the second-order Maxwell operator \(\mathcal M(\omega)\), the positive-semidefinite dissipation operator $\mathcal D(\omega)=-\mathrm{Im}\,\mathcal  M(\omega)
$
contains the complete fluctuation structure of the open problem. 
A Gram factorization of \(\mathcal D(\omega)\) gives a bosonic realization in which medium-assisted (MA) channels, generated by the local imaginary part of the permittivity, and boundary-assisted (BA) channels, generated by the open boundary condition, appear on equal footing. 

The same factorization gives an input-output theory for structured electromagnetic subsystems. 
Operationally, an open boundary condition specifies how the finite electromagnetic (EM) domain exchanges quanta with external DoFs. 
In the Maxwell boundary-value problem, this coupling enters through \(\mathcal D(\omega)\), which identifies the corresponding external reservoir channels and the unmonitored dissipative channels. 
The resulting input-output relation maps incoming external channels to outgoing external channels, with additional noise operators required by material and boundary dissipation. 
This distinguishes it from conventional input-output theory~\cite{gerry2023introductory,PRA_IO_1996,PRA_IO_1999,PRL_IO_2019,Moon2024TAP,moon_IO_v2}, where the external reservoir channels are typically prescribed a priori and often simplified through single-mode or Markov approximations.

To demonstrate the construction, we consider two representative settings. 
First, we calculate the Purcell factor in a dispersive and lossy dielectric structure with mixed outgoing and impedance boundaries.
Second, we apply the same construction to waveguide-port boundary conditions and compute finite-bandwidth Hong--Ou--Mandel (HOM) interference~\cite{PRL_HOM} in two-dimensional photonic integrated circuit (PIC) configurations.

\section{Theory}
Consider the second-order Maxwell equation in the frequency domain,
$\mathcal{M}(\omega)\,\mathbf{E}(\mathbf{r},\omega)
=
i\omega\mu_0\,\mathbf{J}(\mathbf{r},\omega),
\label{eq:maxwell_eq_classical}
$ 
where the Maxwell operator is defined as
$\mathcal{M}(\omega)
=
\nabla \times \left[\mu^{-1}(\mathbf{r},\omega)\,\nabla \times \right]
-\omega^{2}\varepsilon(\mathbf{r},\omega).
\label{eq:maxwell_operator_def}
$
With the $e^{-i\omega t}$ convention, we define the dissipation operator as
$\mathcal{D}(\omega)= -\,\mathrm{Im}\,\mathcal{M}(\omega).
\label{eq:dissipation_operator_def}
$ This operator is nonzero only in the presence of material absorption or radiative leakage induced by open boundary conditions.
Assuming $\mu=1$, the dissipation operator can be decomposed as
$\mathcal{D}(\omega)=\mathcal{D}_{\mathrm{BA}}(\omega)+\mathcal{D}_{\mathrm{MA}}(\omega),
\label{eq:D_split}
$ where $\mathcal{D}_{\mathrm{BA}}$ originates from the curl--curl part of the Maxwell operator and describes BA dissipation, while $\mathcal{D}_{\mathrm{MA}}$ originates from the permittivity term and describes MA dissipation. For passive open systems, both contributions are positive semidefinite and therefore admit channel decompositions,
\begin{equation}
\mathcal{D}_{\mathrm{BA}}(\omega)
=
\int d\alpha\,
\mathbf{C}_{\alpha,\mathrm{BA}}(\omega)\otimes
\mathbf{C}_{\alpha,\mathrm{BA}}^{*}(\omega),
\label{eq:D_BA_factorization}
\end{equation}
and
\begin{equation}
\mathcal{D}_{\mathrm{MA}}(\omega)
=
\int d\beta\,
\mathbf{C}_{\beta,\mathrm{MA}}(\omega)\otimes
\mathbf{C}_{\beta,\mathrm{MA}}^{*}(\omega),
\label{eq:D_MA_factorization}
\end{equation}
which define the BA and MA dissipative channels, respectively.
We now consider the operator-valued quantum Maxwell equation~\cite{Philbin_2010,ciatonni_MLN_2024,Chew_quantum_Maxwell_equation},
$\mathcal{M}(\omega)\,\hat{\mathbf E}(\mathbf r,\omega)
=
i\omega\mu_0\,\hat{\mathbf J}(\mathbf r,\omega)
\label{eq:quantum_maxwell_eq}
$. The field operator can then be written exactly as
\begin{equation}
\hat{\mathbf E}(\mathbf r,\omega)
=
i\omega\mu_0
\int_V d^3r'\,
\mathcal G(\mathbf r,\mathbf r';\omega)\cdot
\hat{\mathbf J}(\mathbf r',\omega).
\label{eq:E_from_J}
\end{equation}
where \(\mathcal G(\mathbf r,\mathbf r';\omega)\) denotes the dyadic Green function.
The standard commutation relation is given by~\cite{Scheel2008MacroscopicQED,ciatonni_MLN_2024}
\begin{equation}
[\hat{\mathbf E}(\mathbf r,\omega),\hat{\mathbf E}^{\dagger}(\mathbf r',\omega')]
=
\frac{\hbar\mu_0\omega^2}{\pi}\,
\mathrm{Im}\,\mathcal G(\mathbf r,\mathbf r';\omega)\,
\delta(\omega-\omega'),
\label{eq:E_commutator}
\end{equation}
together with the Green's identity
\begin{equation}
\begin{aligned}
\mathrm{Im}\,\mathcal G(\mathbf r,\mathbf r';\omega) = & \int_V d^3x\int_V d^3y\, \mathcal G(\mathbf r,\mathbf x;\omega) \\
& \cdot \mathcal D(\mathbf x,\mathbf y;\omega)\cdot \mathcal G^\dagger(\mathbf y,\mathbf r';\omega).
\end{aligned}
\label{eq:green_identity}
\end{equation}
Substituting Eq.~\eqref{eq:E_from_J} into Eq.~\eqref{eq:E_commutator} and comparing with Eq.~\eqref{eq:green_identity}, we obtain
\begin{equation}
[\hat{\mathbf J}(\mathbf r,\omega),\hat{\mathbf J}^{\dagger}(\mathbf r',\omega')]
=
\frac{\hbar}{\pi\mu_0}\,
\mathcal D(\mathbf r,\mathbf r';\omega)\,
\delta(\omega-\omega').
\label{eq:J_commutator}
\end{equation}
Since $\mathcal D$ is given by the channel decompositions in Eqs.~\eqref{eq:D_BA_factorization} and \eqref{eq:D_MA_factorization}, Eq.~\eqref{eq:J_commutator} admits the minimal bosonic realization as
\begin{equation}
\begin{aligned}
\hat{\mathbf J}(\mathbf r,\omega) = & \sqrt{\frac{\hbar}{\pi\mu_0}} \Biggl[ \int d\alpha\, \mathbf C_{\alpha,\mathrm{BA}}(\mathbf r,\omega)\,\hat a_{\alpha,\mathrm{BA}}(\omega) \\
& + \int d\beta\, \mathbf C_{\beta,\mathrm{MA}}(\mathbf r,\omega)\,\hat a_{\beta,\mathrm{MA}}(\omega) \Biggr]
\end{aligned}
\end{equation}
Here, the bosonic operators for each channel satisfy the canonical commutation relations
$[\hat a_{\alpha,\nu}(\omega),\hat a_{\alpha',\nu'}^{\dagger}(\omega')]
=
\delta_{\nu\nu'}\,\delta_{\alpha\alpha'}\,\delta(\omega-\omega'),
\label{eq:channel_boson_comm}
$ where $ \nu,\nu'\in\{\mathrm{BA},\mathrm{MA}\}$. Finally, the quantized field in Eq.~\eqref{eq:E_from_J} can be represented as
\begin{equation}
\begin{aligned}
\hat{\mathbf E}(\mathbf r,\omega) = & \, i\omega\sqrt{\frac{\hbar\mu_0}{\pi}} \int d\nu \\
& \int_V d^3r'\, \mathcal G(\mathbf r,\mathbf r';\omega) \cdot \mathbf C_\nu(\mathbf r',\omega) \, \hat a_\nu(\omega)
\end{aligned}
\label{eq:E_channel_expansion}
\end{equation}

\section{Dissipative Channels}

\begin{table*}[t]
\caption{Representative dissipative channels associated with open boundary conditions. Here $\hat{\mathbf n}$ is the outward unit normal, $\hat{\mathbf e}_{\Omega,s}$ is a transverse polarization basis, $\mathbf e_{t,p,m}$ is the transverse profile of the $m$th propagating mode at port $p$, and $\mathbf Y_s$ is the surface admittance operator. For a general boundary condition, $\lambda_n$ and $\mathbf u_n$ denote the eigenvalues and eigenvectors of $\mathcal D_{\mathrm{BA}}$, whose spectral decomposition gives channels.}
\begin{ruledtabular}
\begin{tabular}{lll}
Boundary Condition & Mathematical Expression& Dissipative Channel \\
\hline
\addlinespace[1.5ex]
Outgoing Boundary Condition
&
$\begin{aligned}
&\mathbf E \sim (e^{ik_0 r}/r)\mathbf F, 
&\nabla\times\mathbf E \sim ik_0 \hat{\mathbf r}\times \mathbf E
\end{aligned}$
&
Equation~\eqref{eq:C_BA_freespace_boundary}

\\[1.0ex]

Impedance Boundary Condition
&
$\displaystyle \hat{\mathbf n}\times\mathbf H
=
\mathbf Y_s\cdot
[\hat{\mathbf n}\times(\hat{\mathbf n}\times\mathbf E)]$
&
$\displaystyle
\mathbf C_{\sigma}(\mathbf r,\omega)
=
\sqrt{\omega\mu_0\,y_\sigma(\omega)}\,
\mathbf e_{\sigma}(\mathbf r_T,\omega)\,
\delta_{\partial V}(\mathbf r)
$

\\[1.0ex]

Port Boundary Condition
&
$\hat{\mathbf n}\times \mathbf H
=
\sum_m a_{p,m}Y_{p,m}\,
\hat{\mathbf n}\times(\hat{\mathbf n}\times \mathbf e_{t,p,m})$
&
$\begin{aligned}
\mathbf C_{p,m} &= \sqrt{\omega\mu_0 \mathrm{Re} \left[Y_{p,m}\right]} \mathbf e_{t,p,m}(\mathbf r_T) \\
&\quad \times \delta(z-z_p), \quad m \in \text{prop}
\end{aligned}$
\\[1.5ex]
General Boundary Condition
&
$\displaystyle
\mathcal D_{\mathrm{BA}}(\omega)
=
\sum_n \lambda_n(\omega)
\mathbf u_n(\omega)\mathbf u_n^\dagger(\omega)
$
&
$\begin{aligned}
\mathbf C_n(\omega)
=
\sqrt{\lambda_n(\omega)}\,\mathbf u_n(\omega)
\end{aligned}$
\end{tabular}
\end{ruledtabular}
\label{tab:BA_channels}
\end{table*}

To identify the dissipative channels explicitly, we use the sesquilinear form associated with the dissipation operator. We write
\begin{equation}
\begin{aligned}
2i\,\langle \mathbf E_1 | \mathcal D(\omega) | \mathbf E_2 \rangle = & \int_V d^3r\, \Bigl[ (\mathcal M(\omega)\mathbf E_1)^*(\mathbf r)\cdot \mathbf E_2(\mathbf r) \\
& - \mathbf E_1^*(\mathbf r)\cdot (\mathcal M(\omega)\mathbf E_2)(\mathbf r) \Bigr]
\end{aligned}
\label{eq:D_sesquilinear}
\end{equation}
Applying the vector Green's identity, we obtain the BA part as
\begin{equation}
\begin{aligned}
2i\,\langle \mathbf E_1 | \mathcal D_{\mathrm{BA}} | \mathbf E_2 \rangle = & \oint_{\partial V} \Bigl[ \mathbf E_1^* \times (\nabla\times \mathbf E_2) \\
& - \mathbf E_2 \times (\nabla\times \mathbf E_1^*) \Bigr] \cdot d\mathbf S
\end{aligned}
\label{eq:D_BA_surface}
\end{equation}
By contrast, the MA part originates from $\mathrm{Im}\,\varepsilon(\mathbf r,\omega)$ and takes the local form
\begin{equation}
\langle \mathbf E_1 | \mathcal D_{\mathrm{MA}} | \mathbf E_2 \rangle
=
\omega^2
\int_V d^3r\,
\mathrm{Im}\,\varepsilon(\mathbf r,\omega)\,
\mathbf E_1^*(\mathbf r)\cdot \mathbf E_2(\mathbf r).
\label{eq:D_MA_form}
\end{equation}
Combining Eqs.~\eqref{eq:D_MA_factorization} and~\eqref{eq:D_MA_form} gives the local MA channels,
\begin{equation}
\mathbf C_{\mathbf r',\xi,\mathrm{MA}}(\mathbf r,\omega)
=
\omega\sqrt{\mathrm{Im}\,\varepsilon(\mathbf r',\omega)}\,
\delta(\mathbf r-\mathbf r')\,\hat{\mathbf e}_\xi,
\label{eq:MA_channel}
\end{equation}
Eqs.~\eqref{eq:MA_channel} and~\eqref{eq:E_channel_expansion} reproduce the standard MQED expression.
Meanwhile, Eq.~\eqref{eq:D_BA_surface} shows that the BA channels are determined by the dissipative geometry imposed by the boundary condition.
As a representative example of the BA construction, we consider the free-space outgoing boundary condition. Specifically, the field satisfies the far-field asymptotics
$\mathbf E(\mathbf r,\omega)\sim \frac{e^{ik_0 r}}{r}\,\mathbf F(\hat{\mathbf r},\omega),
$ and 
$\nabla\times \mathbf E(\mathbf r,\omega)\sim ik_0\,\hat{\mathbf r}\times \mathbf E(\mathbf r,\omega),
$
with $k_0=\omega/c$ and $\hat{\mathbf r}=\mathbf r/r$. Substituting these conditions into Eq.~\eqref{eq:D_BA_surface} and taking $\partial V=S_\infty$, we obtain
\begin{equation}
\langle \mathbf E_1|\mathcal D_{\mathrm{BA}}(\omega)|\mathbf E_2\rangle
=
k_0\int d\Omega\,
\mathbf F_1^*(\hat{\mathbf r},\omega)\cdot
\mathbf F_2(\hat{\mathbf r},\omega).
\label{eq:D_BA_farfield}
\end{equation}
Expanding the far-field amplitude in a transverse polarization basis,
$\mathbf F(\hat{\mathbf r},\omega)
=
\sum_{s=1}^2 F_s(\Omega,\omega)\,\hat{\mathbf e}_{\Omega,s},
$ Eq.~\eqref{eq:D_BA_farfield} becomes
\begin{equation}
\langle \mathbf E_1|\mathcal D_{\mathrm{BA}}(\omega)|\mathbf E_2\rangle
=
k_0\sum_{s=1}^2\int d\Omega\,
F_{1s}^*(\Omega,\omega)\,
F_{2s}(\Omega,\omega).
\label{eq:BA_channels}
\end{equation}
This identifies the free-space BA channels as radiative channels labeled by $(\Omega,s)$. A convenient representation is
\begin{equation}
\begin{aligned}
\mathbf C_{\Omega,s,\mathrm{BA}}(\mathbf r,\omega) = & \, \sqrt{k_0}\lim_{R\to\infty} \Bigl[ 4\pi R\,e^{-ik_0R} \\
& \delta(\mathbf r-R\hat{\mathbf n}_\Omega)\, \hat{\mathbf e}_{\Omega,s} \Bigr],
\end{aligned}
\label{eq:C_BA_freespace_boundary}
\end{equation}
which recovers the MLNF expressions with Eq.~\eqref{eq:E_channel_expansion}. The formulation also provides a consistent route to quasinormal-mode (QNM) quantization~\cite{Franke_PRL_QNMQuantization_2019,QNM_PRA_2024,QNM_PRA_2022,QNM_Arxiv_2025}, where QNM symmetrization is interpreted as a finite-pole projection of the dissipative Maxwell operator.
Table~\ref{tab:BA_channels} summarizes representative BA channels, while the detailed derivations are provided in the Supplemental Material.

\section{Input-Output Theory}
\label{sec:input_output}

The dissipative-channel factorization of the Maxwell operator first gives the field-level quantization of the open electromagnetic system. 
The input-output relation is then obtained by projecting this quantized field onto the boundary-channel basis selected by the measurement. 
An open boundary specifies how the simulated Maxwell domain is coupled to external DoFs, and the output projection extracts the observable channel response selected by that boundary condition. 
This response represents the input-output map selected by the practical boundary condition.

We divide the channel factors into observed external channels \(\mathbf{C}_\mathrm{ext}\) and unobserved channels \(\mathbf{C}_\mathrm{int}\), where the latter include material absorption and boundary leakage that is not explicitly monitored.
For flux-normalized outgoing operators, \(\hat{\mathbf a}_\mathrm{out}(\omega)=\mathbf P_\mathrm{out}(\omega)\hat{\mathbf E}(\omega)\), substituting Eq.~\eqref{eq:E_channel_expansion} gives
\begin{equation}
    \hat{\mathbf a}_\mathrm{out}(\omega)
    =
    \mathbf S(\omega)\hat{\mathbf a}_\mathrm{in}(\omega)
    +
    \mathbf N(\omega)\hat{\mathbf f}_\mathrm{int}(\omega),
\end{equation}
where
\(\mathbf S(\omega)=i\omega\sqrt{\hbar\mu_0/\pi}\,
\mathbf P_\mathrm{out}\mathcal G\mathbf C_\mathrm{ext}\)
and
\(\mathbf N(\omega)=i\omega\sqrt{\hbar\mu_0/\pi}\,
\mathbf P_\mathrm{out}\mathcal G\mathbf C_\mathrm{int}\).
Thus the scattering matrix and its noise kernel are not introduced phenomenologically; both are projections of the same Maxwell Green operator and the same MA/BA channel factors.
This form differs from conventional input-output theory in three essential ways. 
First, \(\mathbf S\) and \(\mathbf N\) retain the full frequency dependence of the electromagnetic structure, so resonances, dispersive materials, and structured boundaries appear as causal memory kernels rather than flat coupling constants.
Second, for the flux-normalized channel basis, the Green identity in Eq.~\eqref{eq:green_identity} enforces
\(\mathbf S(\omega)\mathbf S^\dagger(\omega)+\mathbf N(\omega)\mathbf N^\dagger(\omega)=\mathbf I\).
The noise required to preserve the output commutator is therefore fixed by the Maxwell operator itself, not appended as an external assumption. 
Third, the observed channels can be chosen according to the measurement, while all unobserved material and boundary reservoirs are absorbed into \(\mathbf N(\omega)\). 
The formalism therefore gives a computable input-output relation for realistic open EM devices without requiring a complete mode expansion of the environment.

\section{Numerical Examples}

\begin{figure}
    \centering

    \begin{minipage}{0.95\linewidth}
        \centering
        \includegraphics[width=0.72\linewidth]{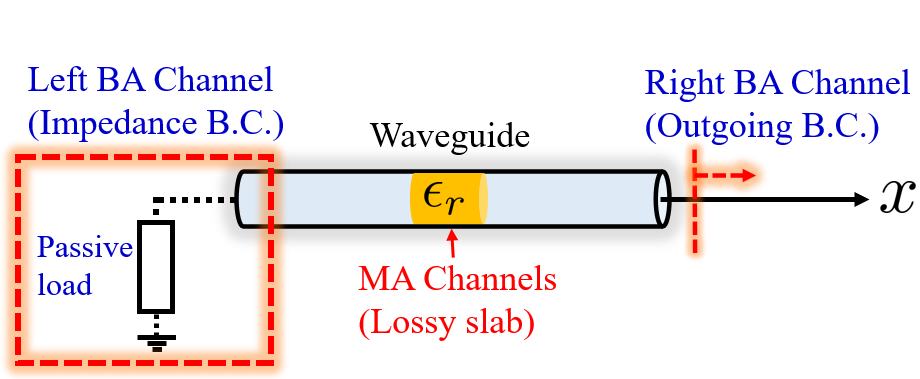}\\[-0.5ex]
        \textbf{(a)}
    \end{minipage}

    \vspace{0.5em}

    \begin{minipage}{0.48\linewidth}
        \centering
        \includegraphics[width=\linewidth]{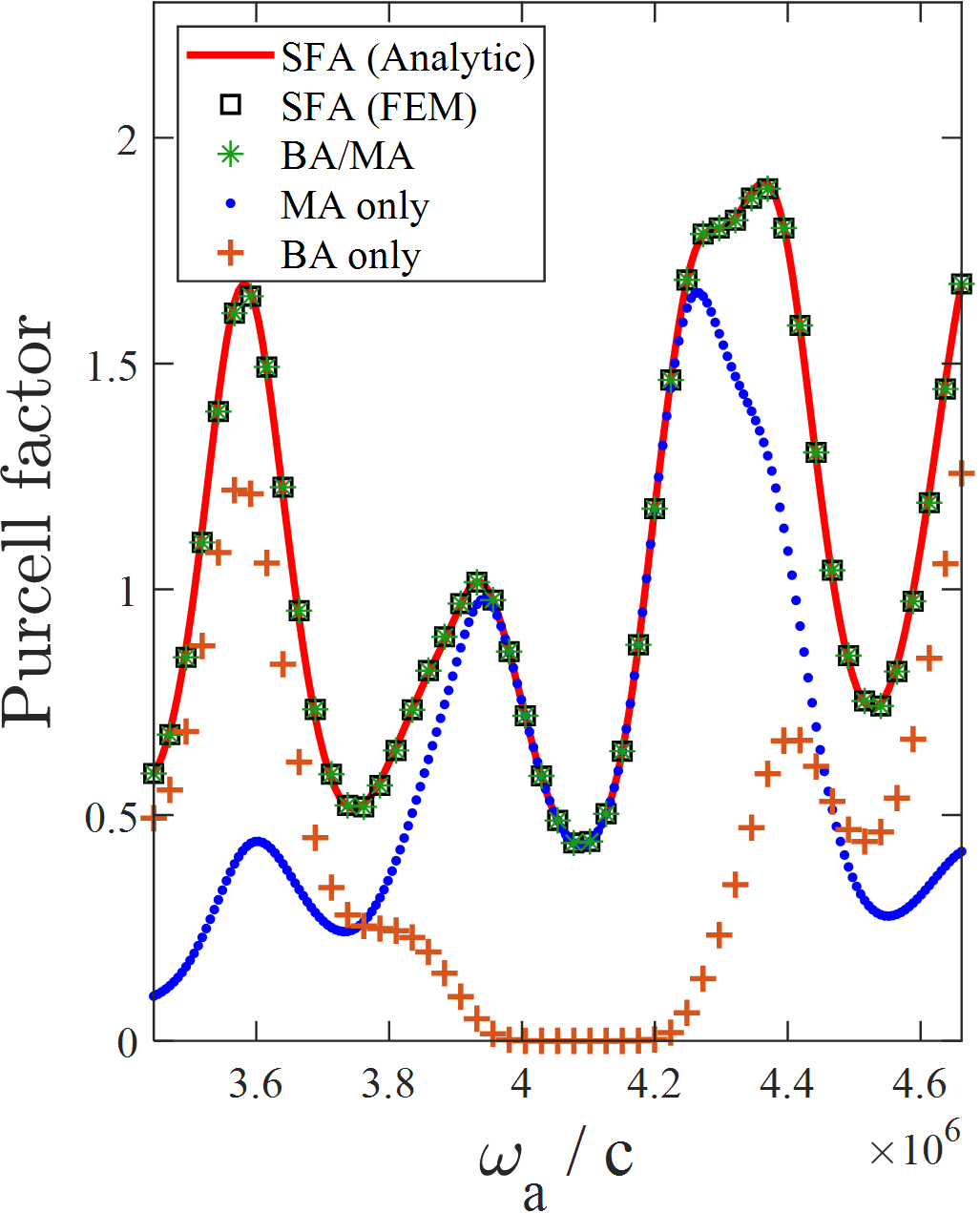}\\[-0.5ex]
        \textbf{(b)}
    \end{minipage}
    \hfill
    \begin{minipage}{0.48\linewidth}
        \centering
        \includegraphics[width=\linewidth]{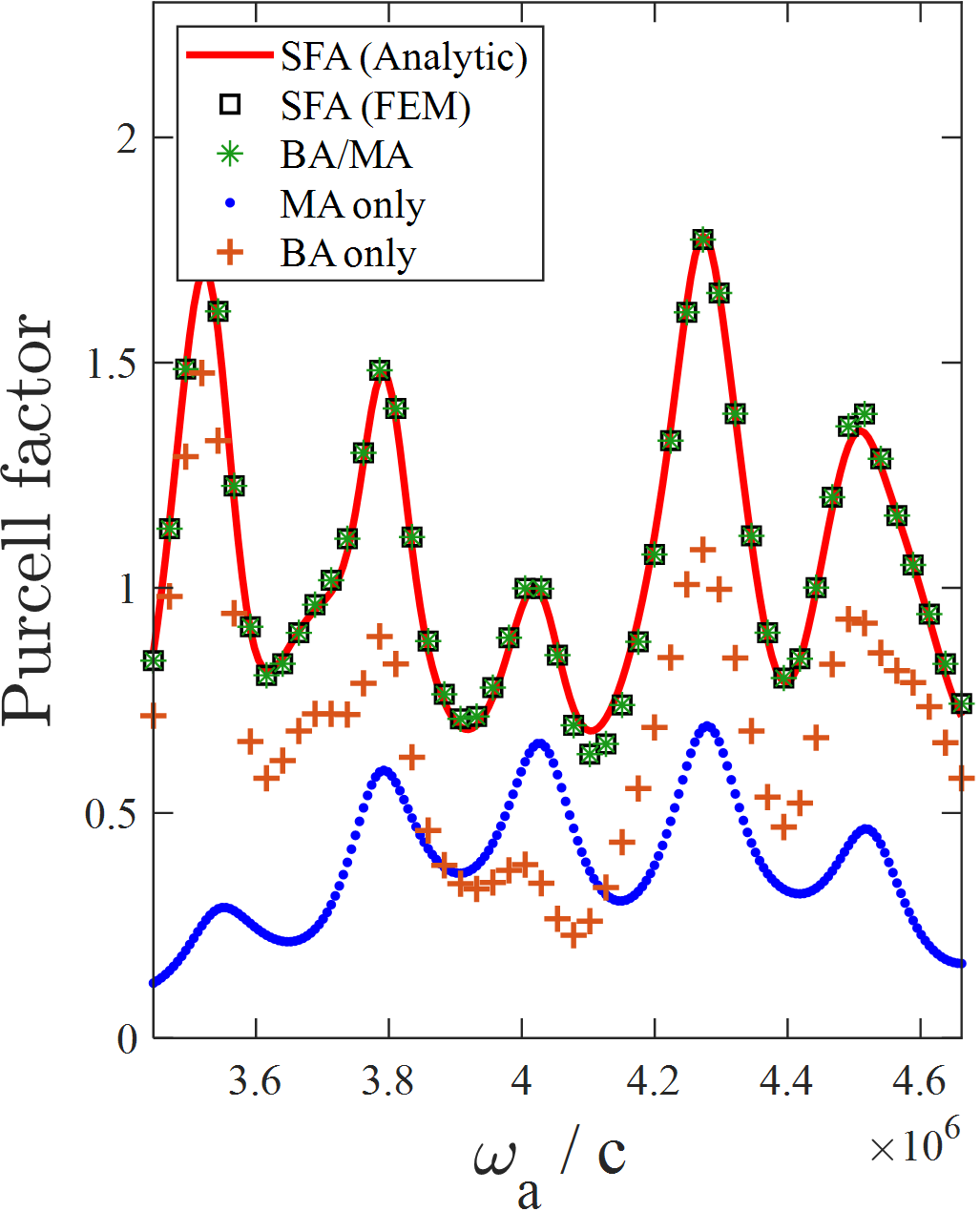}\\[-0.5ex]
        \textbf{(c)}
    \end{minipage}

    \caption{
    One-dimensional waveguide with an impedance and outgoing BA
    channel, with a lossy slab providing MA channels.
    \textbf{(a)} Structure schematic.
    \textbf{(b)} Purcell factor for an emitter inside the slab
    \((x_0 = 5.5\,\mu\text{m})\).
    \textbf{(c)} Purcell factor for an emitter outside the slab
    \((x_0 = 8.5\,\mu\text{m})\).
    }
    \label{fig:IBC_validation}
\end{figure}

We first consider a mixed-boundary system in which material absorption and boundary leakage coexist. 
As shown in Fig.~\ref{fig:IBC_validation}(a), a one-dimensional open waveguide contains a Lorentzian absorptive slab in \(x\in[4.0,7.0]\,\mu{\rm m}\), with resonance wavelength \(\lambda_0=1.55\,\mu{\rm m}\). 
The left end is terminated by a passive impedance load, \(z_L=Z_L/Z_c=1.30+0.70i\), while the right end uses an exact outgoing Dirichlet-to-Neumann boundary condition. 

Figs.~\ref{fig:IBC_validation}(b) and \ref{fig:IBC_validation}(c) show the Purcell factors for emitters inside and outside the lossy slab. 
The full BA-MA reconstruction agrees with both the analytic spectral-function approach and the finite-element Green-function calculation, whereas partial channel reconstructions fail. 
The discrepancy is most visible for an emitter outside the slab, where the local material loss is weak but the electromagnetic fluctuation spectrum is still controlled by boundary leakage. 
Thus the mixed-boundary system is quantized correctly only when both BA and MA dissipative channels are included.

\begin{figure}
    \centering

    \includegraphics[width=\linewidth]{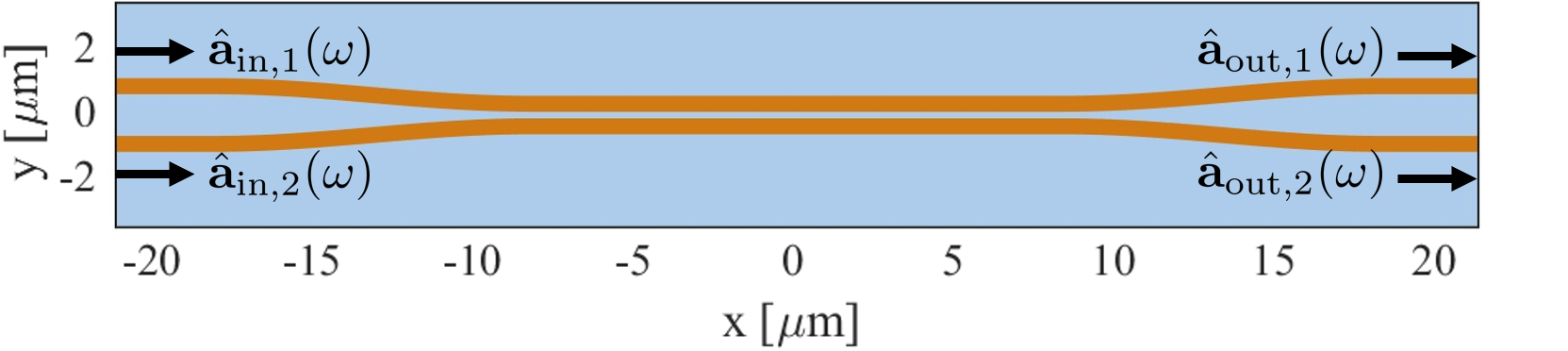}\\[-0.5ex]
    \textbf{(a)}

    \vspace{1.0ex}

    \includegraphics[width=\linewidth]{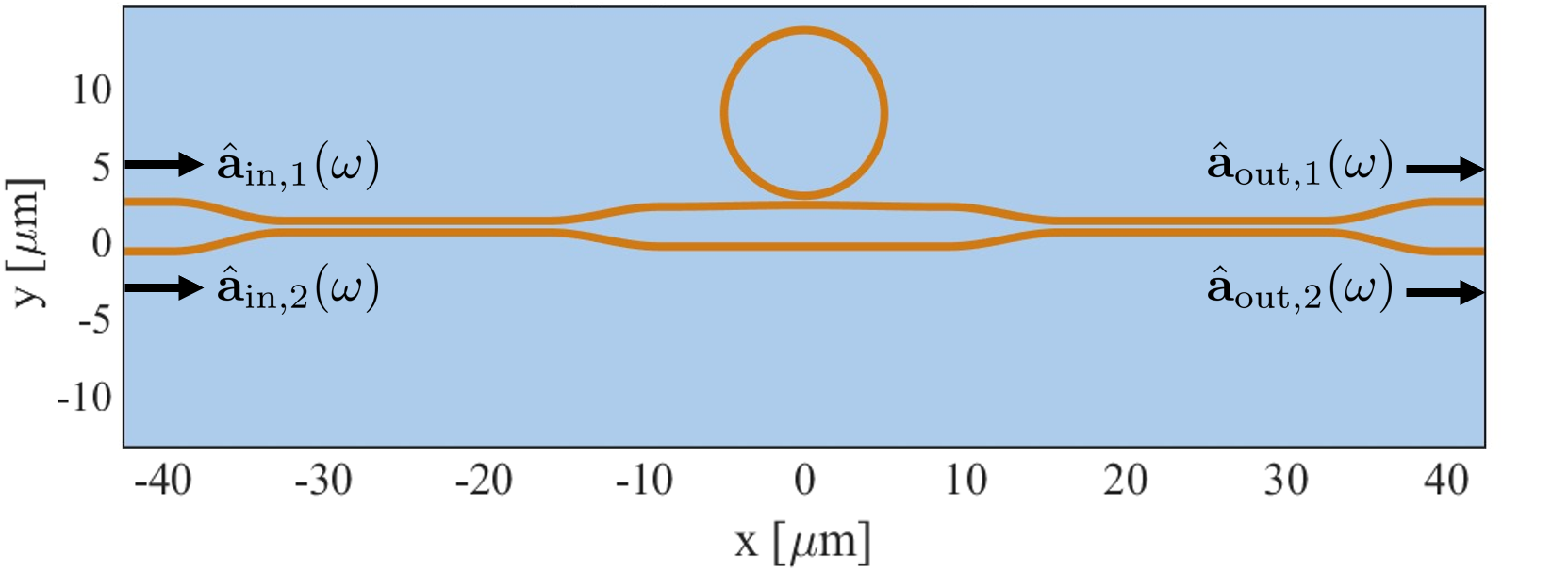}\\[-0.5ex]
    \textbf{(b)}

    \vspace{1.5ex}

    \begin{minipage}{0.49\linewidth}
        \centering
        \includegraphics[width=\linewidth]{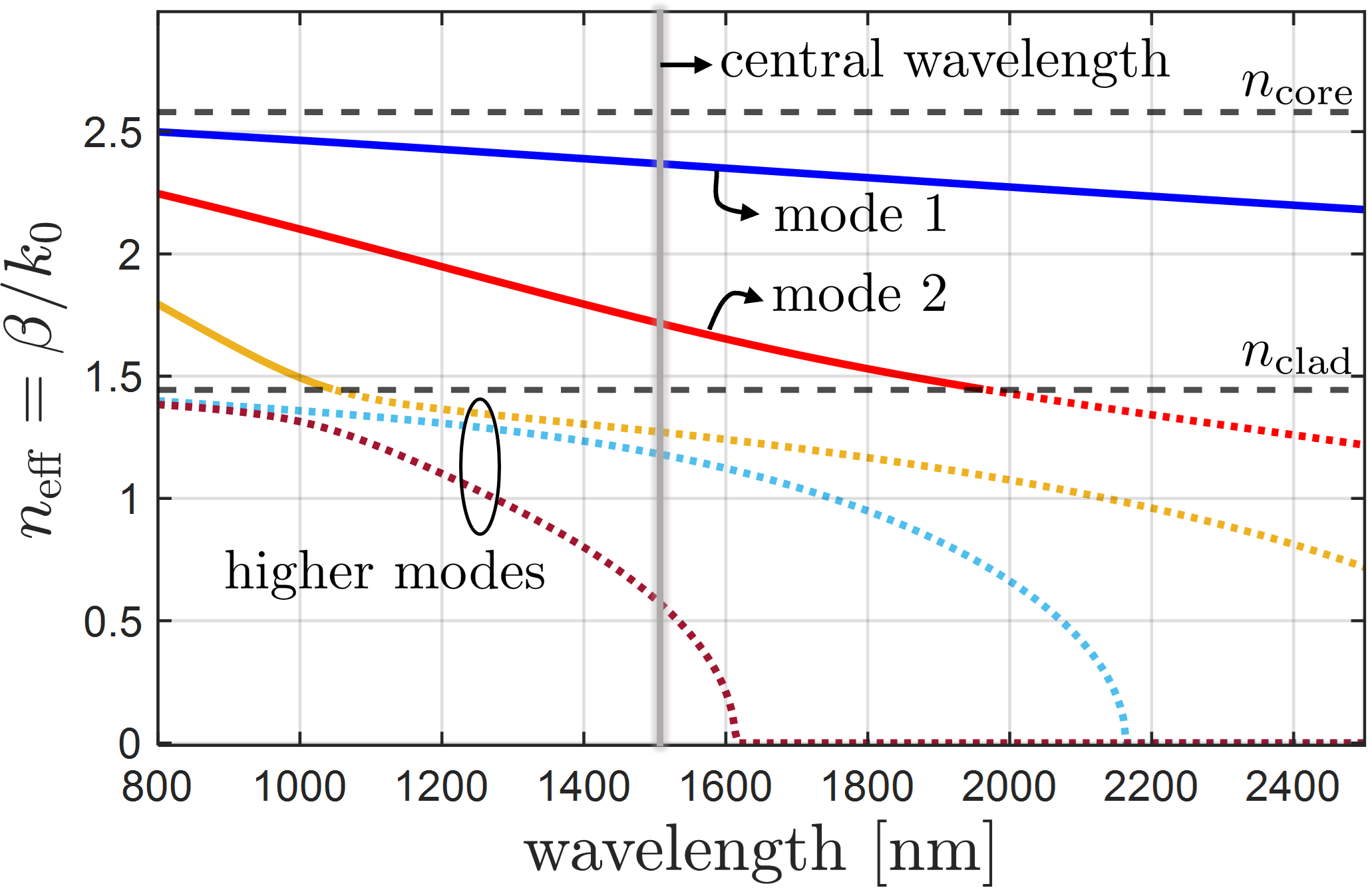}\\[-0.5ex]
        \textbf{(c)}
    \end{minipage}
    \hfill
    \begin{minipage}{0.49\linewidth}
        \centering
        \includegraphics[width=\linewidth]{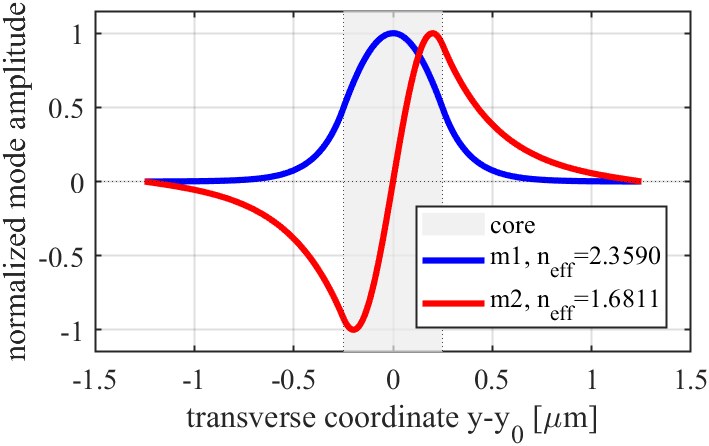}\\[-0.5ex]
        \textbf{(d)}
    \end{minipage}

    \caption{
    Designed photonic integrated circuit configurations and port-channel modes.
    \textbf{(a)} Broadband 50/50 splitter.
    \textbf{(b)} Ring-loaded interferometric splitter.
    \textbf{(c)} Dispersion of the local slab-port cross section, showing the two guided modes retained in the calculation.
    \textbf{(d)} Transverse profiles of the two guided port modes at \(\lambda=1550\,{\rm nm}\).   
    The shaded region denotes the waveguide core.
    }
    \label{fig:port_boundary_modes}
\end{figure}

Next, we use waveguide-port boundaries for the two-dimensional PIC configurations in Fig.~\ref{fig:port_boundary_modes}(a,b). 
The devices are modeled with a scalar effective-index finite-element formulation using a high-index core \(n_{\rm core}=2.58\) and a cladding/background index \(n_{\rm clad}=1.444\). 
Material absorption is included through \(\epsilon''_{\rm core}=2.0\times10^{-4}\) and \(\epsilon''_{\rm clad}=2.0\times10^{-5}\). 
The broadband and ring-loaded devices share the same local slab-port cross section with waveguide width \(w_{\rm wg}=0.50\,\mu{\rm m}\), but are evaluated at their respective center wavelengths, \(\lambda_c^{\rm B}=1550\,{\rm nm}\) and \(\lambda_c^{\rm R}=1557.73\,{\rm nm}\). 
The common slab-port dispersion supports two guided-mode branches throughout this operating wavelength range, while higher-order candidates are below cutoff, as shown in Fig.~\ref{fig:port_boundary_modes}(c). 
The representative transverse profiles of the two retained guided port modes at \(\lambda=1550\,{\rm nm}\) are shown in Fig.~\ref{fig:port_boundary_modes}(d). 
The input photons are injected into the fundamental channels of the two left physical ports, while coincidences are measured by bucket detection over all retained guided channels of the two right physical ports. 
Details of the finite-element implementation, port-channel construction, and remaining simulation parameters are provided in the Supplemental Material.

For each port mode \(\alpha=(p,m)\), where \(p\) labels the physical port and \(m\) labels the guided transverse mode, the mode profile is projected onto the finite-element boundary to define a normalized dual vector \(w_\alpha(\omega)\). 
With the port convention used here, the total scattering kernel, including the prompt contribution, is
\[
S_{\alpha\mu}(\omega)
=
-2i\,\beta_\mu(\omega)
w_\alpha^\dagger(\omega)
\mathbf G_{\rm FEM}(\omega)
w_\mu(\omega)
-
\delta_{\alpha\mu}.
\]
Here \(\mathbf G_{\rm FEM}(\omega)\) is the finite-element Green response, including the two-dimensional device geometry, material absorption, and modal port loading. 
The \(-\delta_{\alpha\mu}\) term is the convention-dependent direct contribution of the incoming channel to the outgoing amplitude.
The finite-bandwidth HOM response is computed by propagating the two single-photon spectra through the full frequency-dependent matrix \(S(\omega)\). 
For comparison, the Markovian reference is obtained from the same lossy device by replacing \(S(\omega)\) with \(S(\omega_c)\). 
We visualize the spectral origin of the interference using the diagnostic kernel
\[
K(\omega)
=
\sum_{a\in\mathcal A}
\sum_{b\in\mathcal B}
\left[
S_{a1}(\omega)S_{b2}(\omega)
+
S_{b1}(\omega)S_{a2}(\omega)
\right],
\]
where \(\mathcal A\) and \(\mathcal B\) denote the two right output ports.

\begin{figure}
    \centering

    \begin{minipage}{0.8\linewidth}
        \centering
        \includegraphics[width=\linewidth]{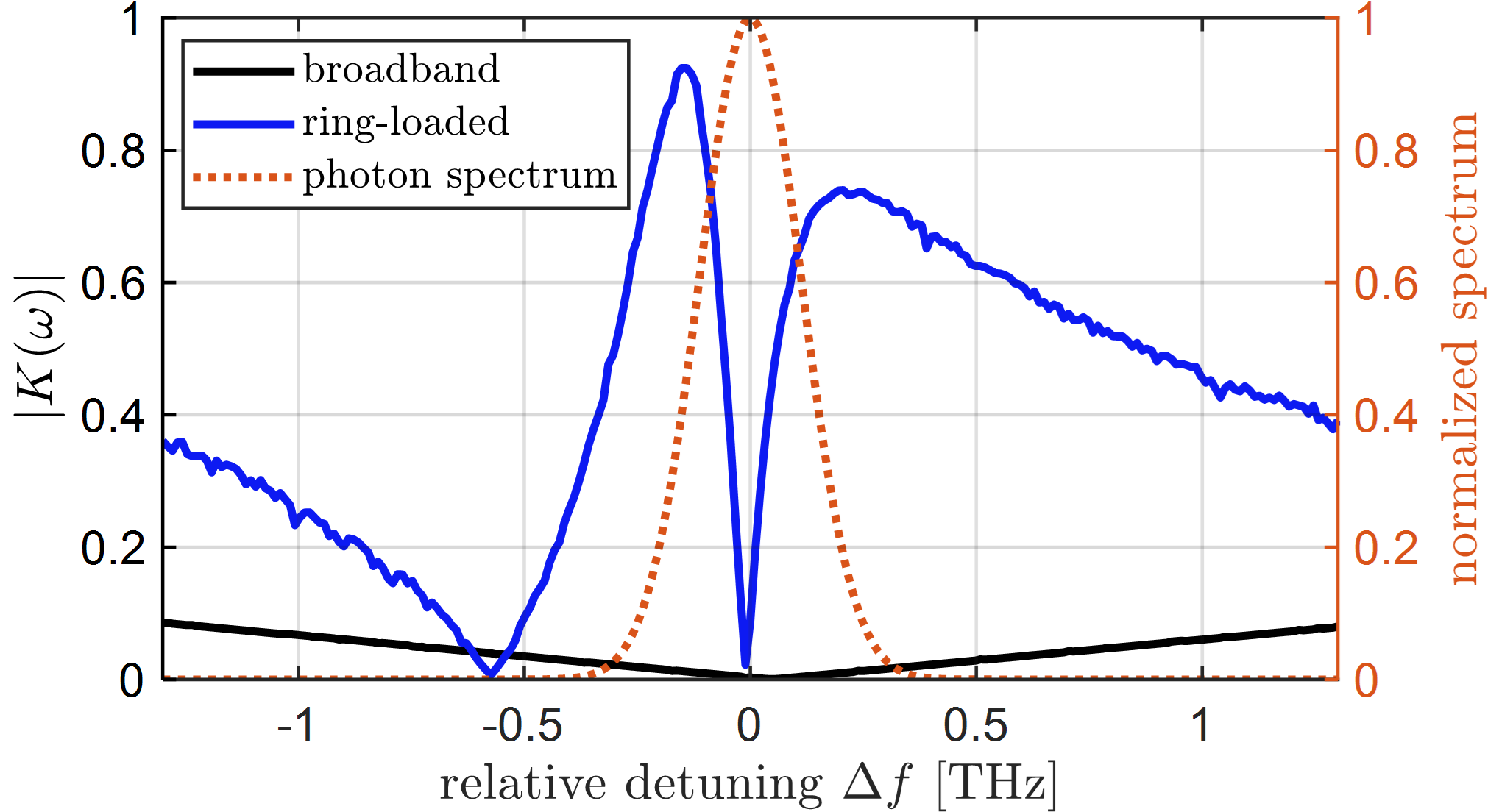}\\[-0.5ex]
        \textbf{(a)}
    \end{minipage}

    \vspace{1.0ex}

    \begin{minipage}{0.8\linewidth}
        \centering
        \includegraphics[width=\linewidth]{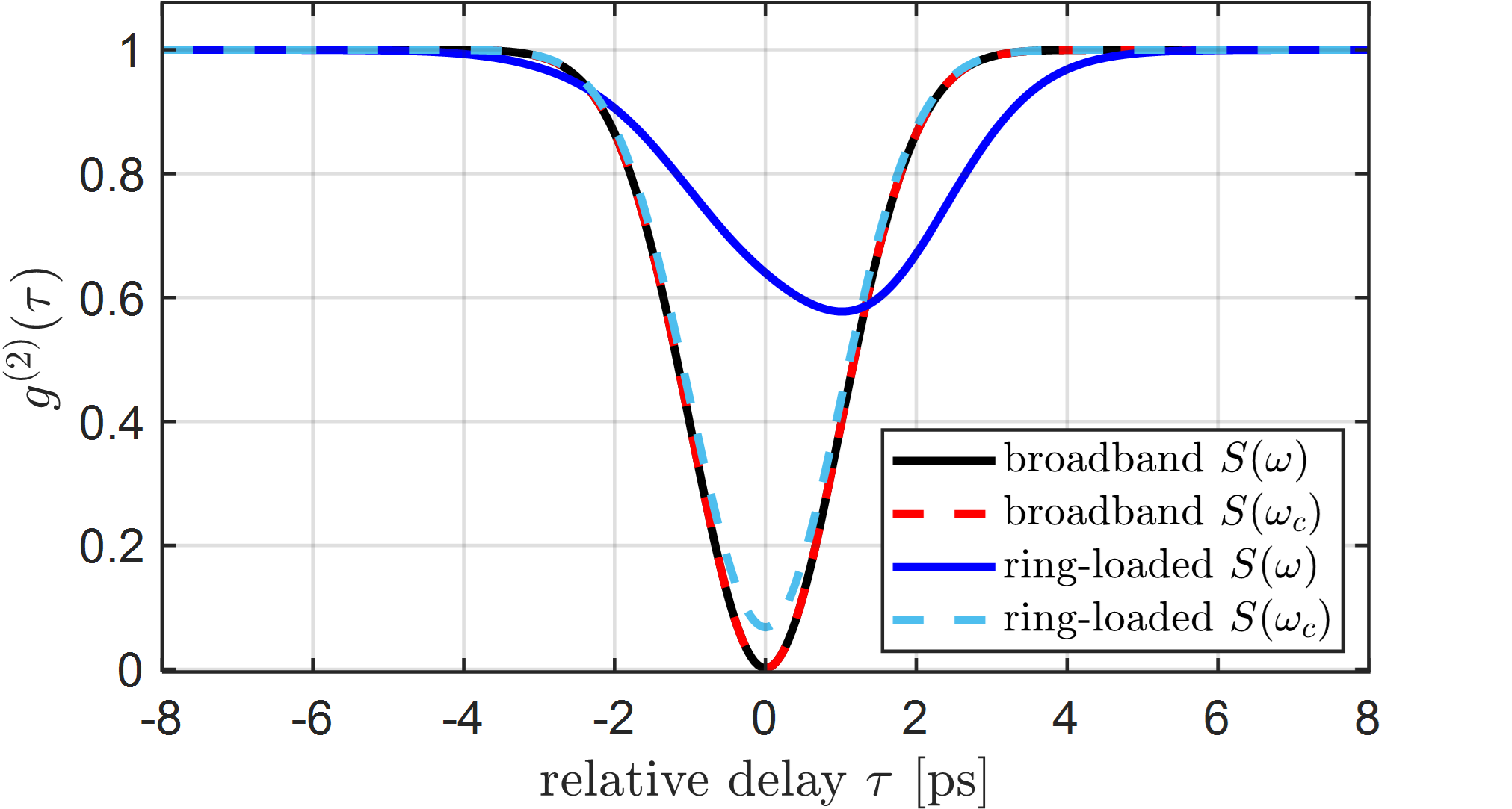}\\[-0.5ex]
        \textbf{(b)}
    \end{minipage}

    \caption{
    HOM interference for non-Markovian splitter responses.
    \textbf{(a)} HOM interference kernel \(K(\omega)\) for the broadband and ring-loaded splitters, with the photon spectrum shown in orange.
    \textbf{(b)} HOM correlation for the full \(S(\omega)\) response and the Markovian flat-\(S(\omega_c)\) reference.
    }
    \label{fig:hom_nonmarkovian_response}
\end{figure}

Figure~\ref{fig:hom_nonmarkovian_response}(a,b) shows the resulting finite-bandwidth response. 
The broadband coupler has an almost flat \(K(\omega)\) over the photon spectrum, so all spectral components undergo nearly the same transformation and the full calculation agrees with the flat-\(S(\omega_c)\) reference. 
In contrast, the ring-loaded splitter is tuned near \(K(\omega_c)\simeq0\), but the ring resonance makes \(K(\omega)\) vary strongly across the photon bandwidth. This produces two finite-bandwidth non-Markovian signatures.
The dip is shallower because different spectral components no longer satisfy the same cancellation condition, leading to incomplete destructive interference after integration over the photon spectrum. 
The dip is also shifted to positive delay because the frequency-dependent scattering phase gives an effective group delay; the external delay that maximizes the two-photon overlap must compensate the temporal memory of the ring. 
Both effects disappear in the Markovian flat-\(S(\omega_c)\) approximation.

\section{Conclusion}

We have presented a dissipative-channel quantization framework for open electromagnetic systems with arbitrary passive boundary conditions.
The central object is the imaginary part of the Maxwell operator, whose channel factorization converts the dissipative part of the Maxwell problem into canonical bosonic reservoirs and yields an exact Green-function expansion of the electric-field operator. 
This construction places medium-assisted fluctuations from material absorption and boundary-assisted fluctuations from environmental exchange on equal footing, including far-field radiation, impedance loads, and guided port channels. 
Finite-element examples verify the Green-identity reconstruction in systems with simultaneous material and boundary dissipation, showing that both MA and BA channels are required to recover the full electromagnetic spectral function. 
The same quantization yields frequency-dependent port input-output kernels and predicts a finite-bandwidth, non-Markovian Hong--Ou--Mandel response in a lossy photonic circuit.
This provides a systematic route to quantum modeling of realistic electromagnetic devices whose reservoirs are fixed by their materials and open boundary conditions.

\bibliography{sorsamp}
\end{document}


\title{\large\bf Supplemental Material for ``Dissipative Channels Determine Open Electromagnetic Quantization''}

\author{Hyunwoo Choi}
\thanks{These authors contributed equally to this work.}
\affiliation{Department of Electrical Engineering, Pohang University of Science and Technology (POSTECH), Pohang 37673, Republic of Korea}

\author{Junwoo Gim}
\thanks{These authors contributed equally to this work.}
\affiliation{Department of Electrical Engineering, Pohang University of Science and Technology (POSTECH), Pohang 37673, Republic of Korea}

\author{Thomas E. Roth}
\affiliation{Elmore Family School of Electrical and Computer Engineering, Purdue University, West Lafayette, Indiana 47907, USA}

\author{Weng Cho Chew}
\affiliation{Elmore Family School of Electrical and Computer Engineering, Purdue University, West Lafayette, Indiana 47907, USA}

\author{Dong-Yeop Na}
\email{dyna22@postech.ac.kr}
\affiliation{Department of Electrical Engineering, Pohang University of Science and Technology (POSTECH), Pohang 37673, Republic of Korea}
\affiliation{Elmore Family School of Electrical and Computer Engineering, Purdue University, West Lafayette, Indiana 47907, USA}
\maketitle
\tableofcontents

\section{Fano-Reservoir Realization of the Channel-Resolved Langevin Current}
\label{app:Fano_reservoir}

In this section, we give a complementary
Fano-reservoir realization of our formulation, paralleling canonical quantization of
macroscopic QED~\cite{Philbin_2010}: the driven reservoir response is absorbed into
the causal Maxwell operator, whereas the homogeneous reservoir solution
remains as the Langevin noise source. In the present formulation, the
local absorptive loss kernel is replaced by the full positive Maxwell dissipation operator $\mathcal D(\omega)
    =
    -\operatorname{Im}\mathcal M(\omega),$
dissipation operator,
whose Gram factors define the reservoir channel couplings.

We consider the Hamiltonian $\hat H
    =
    \hat H_{\rm EM}^{(0)}
    +
    \hat H_{\rm bath}
    +
    \hat H_{\rm int},$
where \(\hat H_{\rm EM}^{(0)}\) generates the lossless reference Maxwell
operator \(\mathcal M_0\). The bath Hamiltonian is
\begin{equation}
    \hat H_{\rm bath}
    =
    \frac{1}{2}
    \int d\nu
    \int_0^\infty d\Omega
    \left[
        \hat \Pi_\nu^\dagger(\Omega)
        \hat \Pi_\nu(\Omega)
        +
        \Omega^2
        \hat X_\nu^\dagger(\Omega)
        \hat X_\nu(\Omega)
    \right].
    \label{eq:Fano_Hbath}
\end{equation}
Here \(\hat X_\nu(\Omega,t)\) is the canonical coordinate of a reservoir
oscillator with bare frequency \(\Omega\), and
\(\hat\Pi_\nu(\Omega,t)\) is its conjugate momentum. They satisfy $    [
    \hat X_\nu(\Omega,t),
    \hat\Pi_{\nu'}^\dagger(\Omega',t)
    ]
    =
    i\hbar
    \delta_{\nu\nu'}\delta(\Omega-\Omega'),$
with all other equal-time commutators vanishing. Equivalently, one may
introduce bosonic operators \(\hat b_\nu(\Omega)\) through 
\begin{equation}
    \hat X_\nu(\Omega)
    =
    \sqrt{\frac{\hbar}{2\Omega}}
    \left[
        \hat b_\nu(\Omega)
        +
        \hat b_\nu^\dagger(\Omega)
    \right],
    \qquad
    \hat\Pi_\nu(\Omega)
    =
    -i
    \sqrt{\frac{\hbar\Omega}{2}}
    \left[
        \hat b_\nu(\Omega)
        -
        \hat b_\nu^\dagger(\Omega)
    \right],
    \label{eq:Fano_XPi_b}
\end{equation}
so that, up to the zero-point energy,
\begin{equation}
    \hat H_{\rm bath}
    =
    \int d\nu
    \int_0^\infty d\Omega\,
    \hbar\Omega\,
    \hat b_\nu^\dagger(\Omega)\hat b_\nu(\Omega).
    \label{eq:Fano_Hbath_b}
\end{equation}
The interaction is chosen so that the reservoir coordinate generates a
generalized polarization coupled linearly to
the electric field,
\begin{equation}
    \hat{\mathbf P}_{\rm R}(\mathbf r,t)
    =
    \int d\nu
    \int_0^\infty d\Omega\,
    \mathbf g_\nu(\mathbf r,\Omega)
    \hat X_\nu(\Omega,t).
    \label{eq:Fano_PR_definition}
\end{equation}
For medium-assisted channels, \(\hat{\mathbf P}_{\rm R}\) is the usual
volume polarization reservoir. For boundary-assisted channels,
\(\hat{\mathbf P}_{\rm R}\) should instead be understood as a
generalized polarization distribution supported on
a dissipative boundary, port, or radiation surface. Equivalently, the
interaction may be written at the Lagrangian level as
\begin{equation}
    \mathcal L_{\rm int}
    =
    \int d^3\mathbf r\,
    \hat{\mathbf P}_{\rm R}(\mathbf r,t)
    \cdot
    \hat{\mathbf E}(\mathbf r,t),
    \label{eq:Fano_Lint}
\end{equation}
where the volume integral denotes the ordinary volume pairing for
medium-assisted channels and the corresponding distributional pairing
for boundary-assisted channels.
The physical reservoir current is the time derivative of this
generalized polarization, $    \hat{\mathbf J}_{\rm R}(\mathbf r,t)
    =
    \partial_t
    \hat{\mathbf P}_{\rm R}(\mathbf r,t).$
With the \(e^{-i\omega t}\) convention, $    \hat{\mathbf J}_{\rm R}(\mathbf r,\omega)
    =
    -i\omega
    \hat{\mathbf P}_{\rm R}(\mathbf r,\omega).$
This relation is the origin of the factor \(-i\omega\) in the
Langevin current.
The Heisenberg equation for the reservoir coordinate has the linear
forced-oscillator form
\begin{equation}
    \left(
        \partial_t^2+\Omega^2
    \right)
    \hat X_\nu(\Omega,t)
    =
    \int d^3\mathbf r\,
    \mathbf g_\nu^\dagger(\mathbf r,\Omega)
    \cdot
    \hat{\mathbf E}(\mathbf r,t).
    \label{eq:Fano_X_EOM_time}
\end{equation}
For boundary-supported channels, the right-hand side is again understood
as the corresponding surface, port-mode, or far-field overlap with the
Maxwell field. In the frequency domain, the retarded solution is
\begin{equation}
    \hat X_\nu(\Omega,\omega)
    =
    \hat X_{\nu,{\rm N}}(\Omega,\omega)
    +
    \frac{
    \int d^3\mathbf r\,
    \mathbf g_\nu^\dagger(\mathbf r,\Omega)
    \cdot
    \hat{\mathbf E}(\mathbf r,\omega)
    }{
    \Omega^2-(\omega+i0^+)^2
    }.
    \label{eq:Fano_X_retarded_solution}
\end{equation}
The first term is the homogeneous reservoir solution and the second term
is the driven response.
Substituting Eq.~\eqref{eq:Fano_X_retarded_solution} into
Eq.~\eqref{eq:Fano_PR_definition} gives
\begin{equation}
    \hat{\mathbf P}_{\rm R}(\mathbf r,\omega)
    =
    \hat{\mathbf P}_{\rm N}(\mathbf r,\omega)
    +
    \int d^3\mathbf r'\,
    \boldsymbol\chi(\mathbf r,\mathbf r';\omega)
    \cdot
    \hat{\mathbf E}(\mathbf r',\omega),
    \label{eq:Fano_P_noise_plus_response}
\end{equation}
where
\begin{equation}
    \hat{\mathbf P}_{\rm N}(\mathbf r,\omega)
    =
    \int d\nu
    \int_0^\infty d\Omega\,
    \mathbf g_\nu(\mathbf r,\Omega)
    \hat X_{\nu,{\rm N}}(\Omega,\omega),
    \label{eq:Fano_PN_definition}
\end{equation}
and
\begin{equation}
    \boldsymbol\chi(\omega)
    =
    \int d\nu
    \int_0^\infty d\Omega\,
    \frac{
    \mathbf g_\nu(\Omega)\otimes
    \mathbf g_\nu^\dagger(\Omega)
    }{
    \Omega^2-(\omega+i0^+)^2
    }.
    \label{eq:Fano_chi_from_g}
\end{equation}
Here the operator notation suppresses spatial coordinates. By construction, the $+i0^+$ regulator enforces strict causality, rendering $\boldsymbol\chi(\omega)$ analytic in the upper half of the complex frequency plane. Consequently, the real and imaginary parts of this generalized susceptibility automatically satisfy the Kramers-Kronig relations. The kernel \(\boldsymbol\chi\) is a generalized susceptibility operator: it may be
volume-supported for material absorption or boundary-supported for
radiation, impedance, waveguide-port, or Floquet-port dissipation.
The reservoir current is therefore
\begin{equation}
    \hat{\mathbf J}_{\rm R}(\omega)
    =
    \hat{\mathbf J}_{\rm N}(\omega)
    -
    i\omega
    \boldsymbol\chi(\omega)
    \hat{\mathbf E}(\omega),
    \label{eq:Fano_J_noise_plus_response}
\end{equation}
with 
\begin{equation}
    \hat{\mathbf J}_{\rm N}(\omega)
    =
    -i\omega
    \hat{\mathbf P}_{\rm N}(\omega).
    \label{eq:Fano_JN_from_PN}
\end{equation}
The deterministic response term is absorbed into the Maxwell operator.
Starting from
\begin{equation}
    \mathcal M_0(\omega)\hat{\mathbf E}(\omega)
    =
    i\omega\mu_0
    \hat{\mathbf J}_{\rm R}(\omega),
    \label{eq:Fano_M0_eq}
\end{equation}
and using Eq.~\eqref{eq:Fano_J_noise_plus_response}, we obtain
\begin{equation}
    \left[
        \mathcal M_0(\omega)
        -
        \omega^2\mu_0
        \boldsymbol\chi(\omega)
    \right]
    \hat{\mathbf E}(\omega)
    =
    i\omega\mu_0
    \hat{\mathbf J}_{\rm N}(\omega).
    \label{eq:Fano_M_chi_eq}
\end{equation}
Thus the macroscopic Maxwell operator generated after eliminating the
driven reservoir response is
\begin{equation}
    \mathcal M(\omega)
    =
    \mathcal M_0(\omega)
    -
    \omega^2\mu_0
    \boldsymbol\chi(\omega).
    \label{eq:Fano_M_from_chi}
\end{equation}
For a lossless reference operator \(\mathcal M_0\), the dissipative part
is
\begin{equation}
    \mathcal D(\omega)
    \equiv
    -
    \operatorname{Im}\mathcal M(\omega)
    =
    \omega^2\mu_0
    \operatorname{Im}
    \boldsymbol\chi(\omega).
    \label{eq:Fano_D_from_chi}
\end{equation}
This identity is understood as an equality of quadratic forms on the
admissible Maxwell field space, allowing both volume-supported and
boundary-supported dissipative operators.
Using
\begin{equation}
    \frac{1}{\Omega^2-(\omega+i0^+)^2}
    =
    \mathcal P
    \frac{1}{\Omega^2-\omega^2}
    +
    i\frac{\pi}{2\omega}
    \delta(\Omega-\omega),
    \qquad
    \omega>0,
    \label{eq:Fano_distribution_identity}
\end{equation}
we find
\begin{equation}
    \operatorname{Im}
    \boldsymbol\chi(\omega)
    =
    \frac{\pi}{2\omega}
    \int d\nu\,
    \mathbf g_\nu(\omega)\otimes
    \mathbf g_\nu^\dagger(\omega).
    \label{eq:Fano_Im_chi}
\end{equation}
Therefore,
\begin{equation}
    \mathcal D(\omega)
    =
    \frac{\pi\omega\mu_0}{2}
    \int d\nu\,
    \mathbf g_\nu(\omega)\otimes
    \mathbf g_\nu^\dagger(\omega).
    \label{eq:Fano_D_from_g}
\end{equation}
Comparing Eq.~\eqref{eq:Fano_D_from_chi} and  Eq.~\eqref{eq:Fano_D_from_g}, the reservoir coupling amplitudes are
\begin{equation}
    \mathbf g_\nu(\mathbf r,\omega)
    =
    \sqrt{
        \frac{2}{\pi\omega\mu_0}
    }\,
    \mathbf C_\nu(\mathbf r,\omega),
    \label{eq:Fano_g_to_C}
\end{equation}
where the channel vectors satisfy the Gram factorization
\begin{equation}
    \mathcal D(\omega)
    =
    \int d\nu\,
    \mathbf C_\nu(\omega)\otimes
    \mathbf C_\nu^\dagger(\omega).
    \label{eq:Fano_D_CC}
\end{equation}
Eqs.~\eqref{eq:Fano_D_from_g} and
\eqref{eq:Fano_g_to_C} reproduce exactly
Eq.~\eqref{eq:Fano_D_CC}. Hence the positive dissipative part of the
Maxwell operator fixes the reservoir spectral density, the channel
index, and the coupling strength.
We emphasize that Eq.~\eqref{eq:Fano_g_to_C} includes both MA and BA
channels. For medium-assisted channels,
\(\mathbf C_{\nu,{\rm MA}}\) is volume-supported and
\(\mathbf g_{\nu,{\rm MA}}\) is an ordinary volume coupling density.
\begin{figure}[t]
    \centering
    \includegraphics[width=0.8\columnwidth]{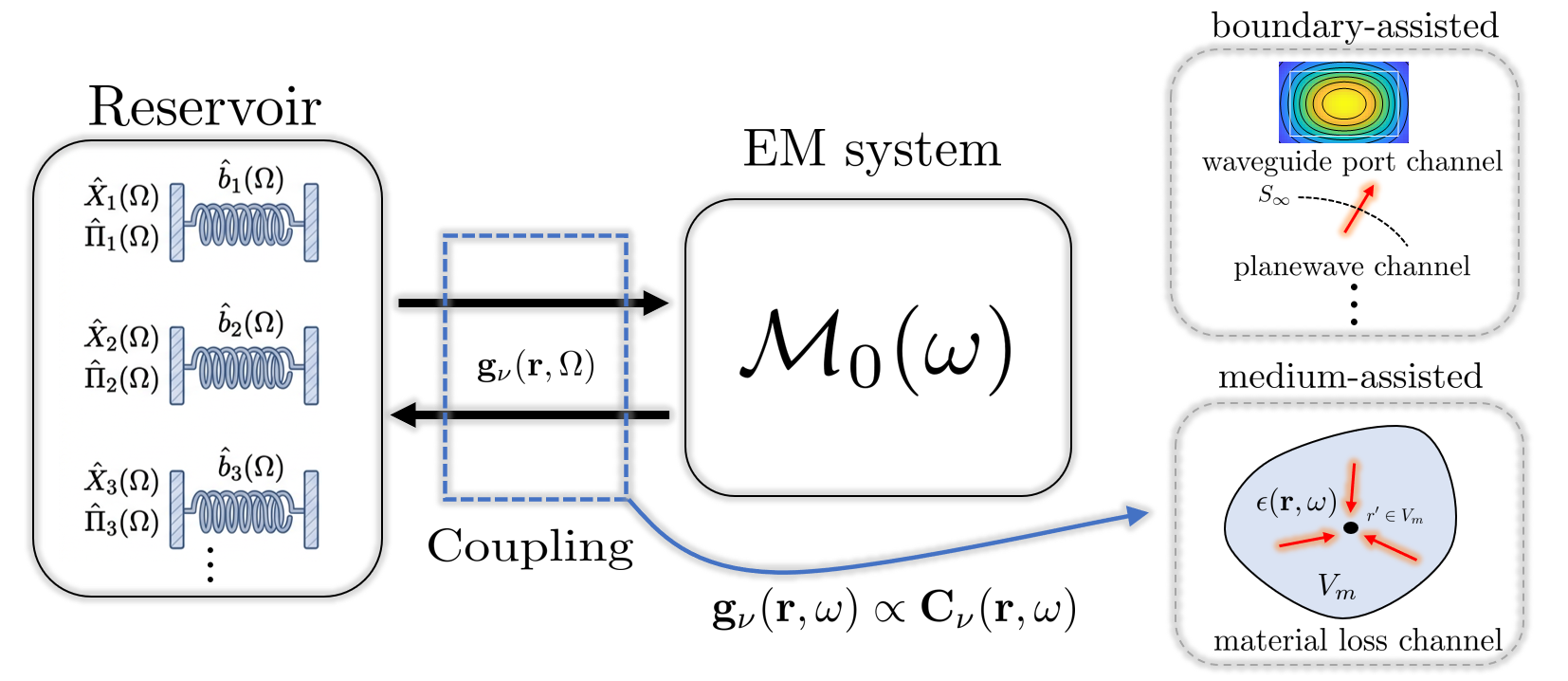}
    \caption{Schematic representation of the Fano-Hopfield reservoir realization bridging Fano/Philbin-type microscopic dynamics and our formulation.}
    \label{fig:Fano_Schematic}
\end{figure}
For boundary-assisted channels,
\begin{equation}
    \mathbf g_{\nu,{\rm BA}}(\mathbf r,\omega)
    =
    \sqrt{
        \frac{2}{\pi\omega\mu_0}
    }\,
    \mathbf C_{\nu,{\rm BA}}(\mathbf r,\omega),
    \label{eq:Fano_g_BA_distribution}
\end{equation}
where \(\mathbf C_{\nu,{\rm BA}}\) may be a boundary-supported
distribution. Examples include surface-admittance channels proportional
to \(\delta_{\partial V}\), waveguide-port channels proportional to
\(\delta_{S_p}\), Floquet-port channels supported on the unit-cell port
surface, and far-field radiation channels obtained as the limiting
surface distribution at infinity. In these cases the pairing $    \int d^3\mathbf r\,
    \mathbf g_{\nu,{\rm BA}}^\dagger(\mathbf r,\omega)
    \cdot
    \hat{\mathbf E}(\mathbf r,\omega)$
denotes the corresponding surface, port-mode, or far-field overlap.
Thus the Fano reservoir is not restricted to a bulk material
polarization bath. Boundary-assisted reservoirs are realized as
auxiliary oscillator continua coupled to the Maxwell field through
boundary-supported generalized polarization
coordinates.
Equivalently, the BA part of the response kernel is the
boundary-supported part of the same generalized susceptibility. Its
absorptive contribution gives
\begin{equation}
    \mathcal D_{\rm BA}(\omega)
    =
    \int d\nu_{\rm BA}\,
    \mathbf C_{\nu,{\rm BA}}(\omega)\otimes
    \mathbf C_{\nu,{\rm BA}}^\dagger(\omega),
    \label{eq:Fano_D_BA_distributional}
\end{equation}
as an identity of quadratic forms. The MA part is obtained analogously
from the volume-supported channels,
\begin{equation}
    \mathcal D_{\rm MA}(\omega)
    =
    \int d\nu_{\rm MA}\,
    \mathbf C_{\nu,{\rm MA}}(\omega)\otimes
    \mathbf C_{\nu,{\rm MA}}^\dagger(\omega).
    \label{eq:Fano_D_MA_distributional}
\end{equation}
Consequently, the same Fano realization covers both material absorption and
open-boundary radiation or leakage.
Finally, the positive-frequency homogeneous solution may be written as
\begin{equation}
    \hat X_{\nu,{\rm N}}^{(+)}(\Omega,\omega)
    =
    \sqrt{\frac{\hbar}{2\Omega}}\,
    \hat b_\nu(\Omega)
    \delta(\omega-\Omega),
    \label{eq:Fano_XN_positive}
\end{equation}
with $    [
    \hat b_\nu(\omega),
    \hat b_{\nu'}^\dagger(\omega')
    ]
    =
    \delta_{\nu\nu'}\delta(\omega-\omega').$
Using Eqs.~\eqref{eq:Fano_PN_definition},
\eqref{eq:Fano_JN_from_PN}, and \eqref{eq:Fano_g_to_C}, we obtain for
\(\omega>0\)
\begin{equation}
    \hat{\mathbf J}_{\rm N}^{(+)}(\mathbf r,\omega)
    =
    -i
    \sqrt{
        \frac{\hbar}{\pi\mu_0}
    }
    \int d\nu\,
    \mathbf C_\nu(\mathbf r,\omega)
    \hat b_\nu(\omega).
    \label{eq:Fano_JN_final_with_phase}
\end{equation}
The overall phase \(-i\) can be absorbed into the definition of the
bosonic reservoir operator. Therefore,
\begin{equation}
    \hat{\mathbf J}_{\rm N}(\mathbf r,\omega)
    =
    \sqrt{
        \frac{\hbar}{\pi\mu_0}
    }
    \int d\nu\,
    \mathbf C_\nu(\mathbf r,\omega)
    \hat b_\nu(\omega),
    \label{eq:Fano_JN_final}
\end{equation}
which is the channel-resolved Langevin current used in the main text.

\section{Fluctuation-Dissipation Relation}
\label{app:FDT_finiteT}

We now extend the standard finite-temperature fluctuation-dissipation
structure of macroscopic QED~\cite{Scheel2008MacroscopicQED} to the present dissipative-channel
formulation. In conventional macroscopic QED, the thermal fluctuation
relations are written in terms of the local absorptive response.

This extension assumes global thermal equilibrium, namely that all
independent dissipative channels are thermalized at the same temperature
\(T\). Using the composite channel label \(\nu\) to denote the full set
of medium-assisted and boundary-assisted channels, we assume $\langle
\hat a_\nu^\dagger(\omega)
\hat a_{\nu'}(\omega')
\rangle_T
=
n_T(\omega)\,
\delta_{\nu\nu'}\delta(\omega-\omega'),$ and $\langle
\hat a_\nu(\omega)
\hat a_{\nu'}^\dagger(\omega')
\rangle_T
=
\big[n_T(\omega)+1\big]\,
\delta_{\nu\nu'}\delta(\omega-\omega'),$
where $    n_T(\omega)
    =
    \frac{1}{e^{\hbar\omega/(k_B T)}-1}.$
Using the channel-resolved Langevin current
\begin{equation}
    \hat{\mathbf J}(\mathbf r,\omega)
    =
    \sqrt{\frac{\hbar}{\pi\mu_0}}
    \int d\nu\,
    \mathbf C_\nu(\mathbf r,\omega)\hat a_\nu(\omega),
    \label{eq:FDT_J_channel}
\end{equation}
together with the Gram factorization of the full dissipation operator,
\begin{equation}
    \mathcal D(\mathbf r,\mathbf r';\omega)
    =
    \int d\nu\,
    \mathbf C_\nu(\mathbf r,\omega)
    \otimes
    \mathbf C_\nu^*(\mathbf r',\omega),
    \label{eq:FDT_D_Gram}
\end{equation}
the unsymmetrized thermal current correlations become
\begin{align}
\langle
\hat{\mathbf J}(\mathbf r,\omega)
\hat{\mathbf J}^\dagger(\mathbf r',\omega')
\rangle_T
&=
\frac{\hbar}{\pi\mu_0}
\big[n_T(\omega)+1\big]\,
\mathcal D(\mathbf r,\mathbf r';\omega)\,
\delta(\omega-\omega'),
\label{eq:FDT_JJdag}
\\
\langle
\hat{\mathbf J}^\dagger(\mathbf r',\omega')
\hat{\mathbf J}(\mathbf r,\omega)
\rangle_T
&=
\frac{\hbar}{\pi\mu_0}
n_T(\omega)\,
\mathcal D(\mathbf r,\mathbf r';\omega)\,
\delta(\omega-\omega').
\label{eq:FDT_JdagJ}
\end{align}
Their difference reproduces the temperature-independent source
commutator, while their sum yields the symmetrized current fluctuation,
\begin{equation}
\left\langle
\left\{
\hat{\mathbf J}(\mathbf r,\omega),
\hat{\mathbf J}^\dagger(\mathbf r',\omega')
\right\}
\right\rangle_T
=
\frac{\hbar}{\pi\mu_0}
\coth\!\left(\frac{\hbar\omega}{2k_B T}\right)
\mathcal D(\mathbf r,\mathbf r';\omega)\,
\delta(\omega-\omega').
\label{eq:FDT_J_sym}
\end{equation}
We now propagate these fluctuations to the electric field operator,
\begin{equation}
\hat{\mathbf E}(\mathbf r,\omega)
=
i\omega\mu_0
\int_V d^3\mathbf x\,
\mathcal G(\mathbf r,\mathbf x;\omega)
\cdot
\hat{\mathbf J}(\mathbf x,\omega).
\label{eq:FDT_E_from_J_repeat}
\end{equation}
Substituting Eq.~\eqref{eq:FDT_JJdag} into
Eq.~\eqref{eq:FDT_E_from_J_repeat}, we obtain
\begin{align}
\langle
\hat{\mathbf E}(\mathbf r,\omega)
\hat{\mathbf E}^\dagger(\mathbf r',\omega')
\rangle_T
&=
\frac{\hbar\mu_0\omega^2}{\pi}
\big[n_T(\omega)+1\big]\,
\delta(\omega-\omega')
\nonumber\\
&\quad\times
\int_V d^3\mathbf x
\int_V d^3\mathbf y\,
\mathcal G(\mathbf r,\mathbf x;\omega)
\cdot
\mathcal D(\mathbf x,\mathbf y;\omega)
\cdot
\mathcal G^\dagger(\mathbf y,\mathbf r';\omega).
\label{eq:FDT_EEdag_pre}
\end{align}
Using the Green spectral identity $    \operatorname{Im}\mathcal G(\omega)
    =
    \mathcal G(\omega)\mathcal D(\omega)\mathcal G^\dagger(\omega),$
this becomes
\begin{equation}
\langle
\hat{\mathbf E}(\mathbf r,\omega)
\hat{\mathbf E}^\dagger(\mathbf r',\omega')
\rangle_T
=
\frac{\hbar\mu_0\omega^2}{\pi}
\big[n_T(\omega)+1\big]\,
\operatorname{Im}\mathcal G(\mathbf r,\mathbf r';\omega)\,
\delta(\omega-\omega').
\label{eq:FDT_EEdag}
\end{equation}
Similarly,
\begin{equation}
\langle
\hat{\mathbf E}^\dagger(\mathbf r',\omega')
\hat{\mathbf E}(\mathbf r,\omega)
\rangle_T
=
\frac{\hbar\mu_0\omega^2}{\pi}
n_T(\omega)\,
\operatorname{Im}\mathcal G(\mathbf r,\mathbf r';\omega)\,
\delta(\omega-\omega').
\label{eq:FDT_EdagE}
\end{equation}
Therefore, the electric-field anticommutator fluctuation is
\begin{equation}
\left\langle
\left\{
\hat{\mathbf E}(\mathbf r,\omega),
\hat{\mathbf E}^\dagger(\mathbf r',\omega')
\right\}
\right\rangle_T
=
\frac{\hbar\mu_0\omega^2}{\pi}
\coth\!\left(\frac{\hbar\omega}{2k_B T}\right)
\operatorname{Im}\mathcal G(\mathbf r,\mathbf r';\omega)\,
\delta(\omega-\omega').
\label{eq:FDT_E_sym}
\end{equation}
Eqs.~\eqref{eq:FDT_J_sym} and~\eqref{eq:FDT_E_sym} are the
finite-temperature equilibrium fluctuation-dissipation relations in the
present channel-resolved formulation. They show that the usual
macroscopic-QED thermal fluctuation structure extends directly once the
local material-loss kernel is replaced by the full dissipative operator
\(\mathcal D(\omega)\).
Under global thermal equilibrium, the same equilibrium FDT therefore
applies to medium-assisted and boundary-assisted dissipative channels,
including material absorption, radiative leakage, waveguide-port
channels, and impedance-boundary contributions. For nonequilibrium
situations, such as different ports or radiation channels held at
different temperatures, the individual channel occupations should be
specified separately rather than using a single Bose factor
\(n_T(\omega)\).

\section{Boundary-Assisted Channel Expressions}
\label{app:BA_channels}

In this appendix, we derive representative boundary-assisted (BA)
dissipative channels associated with impedance, waveguide-port,
Floquet-port, and general passive numerical boundary conditions.
Throughout, we use the sesquilinear form of the BA dissipation operator
introduced in the main text,
\begin{equation}
2i
\langle \mathbf E_1 |
\mathcal D_{\mathrm{BA}}(\omega)
| \mathbf E_2 \rangle
=
\oint_{\partial V}
\left[
\mathbf E_1^* \times (\nabla\times \mathbf E_2)
-
\mathbf E_2 \times (\nabla\times \mathbf E_1^*)
\right]\cdot d\mathbf S .
\label{eq:app_BA_surface_start}
\end{equation}
The outward surface normal is used throughout, and the sign convention is
chosen such that an outgoing passive boundary gives
\(\langle \mathbf E|\mathcal D_{\mathrm{BA}}|\mathbf E\rangle\ge 0\).
With the \(e^{-i\omega t}\) convention and, for simplicity, a
nonmagnetic exterior medium, we use $    \nabla\times \mathbf E = i\omega\mu_0 \mathbf H .
$
Then Eq.~\eqref{eq:app_BA_surface_start} can be rewritten in terms of the
tangential boundary fields. The resulting quadratic form identifies the positive-semidefinite BA
operator and hence the corresponding Gram-factorized channel vectors.

\subsection{Impedance boundary condition}

We consider an impedance boundary condition on a surface \(\partial V\)
of the form
\begin{equation}
\hat{\mathbf n}\times \mathbf H
=
\mathbf Y_s(\omega)\cdot
\left[
\hat{\mathbf n}\times
(\hat{\mathbf n}\times \mathbf E)
\right].
\label{eq:app_impedance_bc}
\end{equation}
Equivalently, if $    \mathbf E_T
    =
    \mathbf E
    -
    \hat{\mathbf n}(\hat{\mathbf n}\cdot\mathbf E)$
denotes the tangential electric field, then
\begin{equation}
    \hat{\mathbf n}\times \mathbf H
    =
    -\mathbf Y_s(\omega)\cdot \mathbf E_T .
\end{equation}
Here \(\mathbf Y_s\) is the surface admittance operator acting on
tangential fields. With the outward-normal convention used above, a
passive impedance boundary satisfies $    \operatorname{Re}\mathbf Y_s
    \equiv
    \frac{\mathbf Y_s+\mathbf Y_s^\dagger}{2}
    \ge 0$
on the tangential surface Hilbert space.

Substituting Eq.~\eqref{eq:app_impedance_bc} into
Eq.~\eqref{eq:app_BA_surface_start} yields
\begin{equation}
\langle \mathbf E_1 |
\mathcal D_{\mathrm{BA}}^{(\mathrm{imp})}(\omega)
| \mathbf E_2 \rangle
=
\omega\mu_0
\oint_{\partial V} dS\;
\mathbf E_{1,T}^*(\mathbf r_T,\omega)\cdot
\operatorname{Re}\mathbf Y_s(\omega)\cdot
\mathbf E_{2,T}(\mathbf r_T,\omega).
\label{eq:app_impedance_quadratic}
\end{equation}
Since \(\operatorname{Re}\mathbf Y_s\) is positive semidefinite for a
passive boundary, it admits the spectral representation
\begin{equation}
\operatorname{Re}\mathbf Y_s(\omega)
=
\sum_\sigma
y_\sigma(\omega)\,
\mathbf e_\sigma(\omega)
\otimes
\mathbf e_\sigma^*(\omega),
\qquad
y_\sigma(\omega)\ge 0 ,
\label{eq:app_impedance_spectral}
\end{equation}
where \(\{\mathbf e_\sigma\}\) is an orthonormal tangential basis on the
boundary. In kernel notation, the surface-coordinate dependence of
\(\mathbf e_\sigma\) is understood.

Using Eq.~\eqref{eq:app_impedance_spectral},
Eq.~\eqref{eq:app_impedance_quadratic} becomes
\begin{equation}
\langle \mathbf E_1 |
\mathcal D_{\mathrm{BA}}^{(\mathrm{imp})}(\omega)
| \mathbf E_2 \rangle
=
\omega\mu_0
\sum_\sigma
y_\sigma(\omega)\,
b_{1,\sigma}^*(\omega)b_{2,\sigma}(\omega),
\label{eq:app_impedance_modal_quadratic}
\end{equation}
with
\begin{equation}
b_{\ell,\sigma}(\omega)
=
\oint_{\partial V} dS\;
\mathbf e_\sigma^*(\mathbf r_T,\omega)\cdot
\mathbf E_{\ell,T}(\mathbf r_T,\omega).
\label{eq:app_impedance_coeff}
\end{equation}
Therefore,
\begin{equation}
\mathcal D_{\mathrm{BA}}^{(\mathrm{imp})}(\omega)
=
\sum_\sigma
\mathbf C_\sigma(\omega)
\otimes
\mathbf C_\sigma^*(\omega),
\label{eq:app_impedance_gram}
\end{equation}
with impedance-boundary channels
\begin{equation}
\mathbf C_\sigma(\mathbf r,\omega)
=
\sqrt{\omega\mu_0\,y_\sigma(\omega)}\;
\mathbf e_\sigma(\mathbf r_T,\omega)\,
\delta_{\partial V}(\mathbf r).
\label{eq:app_impedance_channel_final}
\end{equation}
The surface delta distribution is defined by
\begin{equation}
\int_V d^3\mathbf r\;
\delta_{\partial V}(\mathbf r)f(\mathbf r)
=
\oint_{\partial V} dS\; f|_{\partial V}.
\label{eq:app_surface_delta}
\end{equation}

\subsection{Waveguide-port boundary condition}

Consider a port cross section \(S_p\) located at \(z=z_p\), with outward
normal \(\hat{\mathbf n}\) and transverse coordinate \(\mathbf r_T\).
We expand the tangential electric field on the port in a transverse modal
basis as
\begin{equation}
\mathbf E_{T,\ell}^{(p)}(\mathbf r_T,\omega)
=
\sum_m
a_{\ell,p,m}(\omega)
\mathbf e_{t,p,m}(\mathbf r_T,\omega),
\qquad
\ell=1,2.
\label{eq:app_port_E_expand}
\end{equation}
The outgoing port boundary condition is written as
\begin{equation}
\hat{\mathbf n}\times \mathbf H_\ell
=
\sum_m
a_{\ell,p,m}(\omega)
Y_{p,m}(\omega)\,
\hat{\mathbf n}\times
\left[
\hat{\mathbf n}\times \mathbf e_{t,p,m}
\right].
\label{eq:app_port_bc}
\end{equation}
Since
\(\hat{\mathbf n}\times(\hat{\mathbf n}\times\mathbf e_{t,p,m})
=-\mathbf e_{t,p,m}\),
this convention is consistent with the impedance-boundary convention
used above. The modal admittance \(Y_{p,m}\) is defined so that passive
propagating outgoing modes satisfy
\(\operatorname{Re}Y_{p,m}>0\).
We assume the transverse modes are orthonormal on the port cross section,
\begin{equation}
\int_{S_p} d^2\mathbf r_T\,
\mathbf e_{t,p,m}^*(\mathbf r_T,\omega)\cdot
\mathbf e_{t,p,n}(\mathbf r_T,\omega)
=
\delta_{mn}.
\label{eq:app_port_orth}
\end{equation}
Substituting Eq.~\eqref{eq:app_port_bc} into
Eq.~\eqref{eq:app_BA_surface_start}, restricting the surface integral to
the port section, and using Eq.~\eqref{eq:app_port_orth}, we obtain
\begin{equation}
\langle \mathbf E_1 |
\mathcal D_{\mathrm{BA}}^{(p)}(\omega)
| \mathbf E_2 \rangle
=
\omega\mu_0
\sum_m
\operatorname{Re}Y_{p,m}(\omega)\,
a_{1,p,m}^*(\omega)a_{2,p,m}(\omega).
\label{eq:app_port_quadratic}
\end{equation}
Only propagating modes contribute to the dissipative part. For evanescent
modes the corresponding modal admittance is purely imaginary and hence
\(\operatorname{Re}Y_{p,m}=0\). Therefore,
\begin{equation}
\langle \mathbf E_1 |
\mathcal D_{\mathrm{BA}}^{(p)}(\omega)
| \mathbf E_2 \rangle
=
\sum_{m\in\mathrm{prop}}
\left\langle
\mathbf E_1
\middle|
\mathbf C_{p,m}(\omega)
\otimes
\mathbf C_{p,m}^*(\omega)
\middle|
\mathbf E_2
\right\rangle ,
\label{eq:app_port_gram_form}
\end{equation}
with port-channel vectors
\begin{equation}
\mathbf C_{p,m}(\mathbf r,\omega)
=
\sqrt{\omega\mu_0\,\operatorname{Re}Y_{p,m}(\omega)}\;
\mathbf e_{t,p,m}(\mathbf r_T,\omega)\,
\delta_{S_p}(\mathbf r),
\qquad
m\in\mathrm{prop}.
\label{eq:app_port_channel_final}
\end{equation}
Here \(\delta_{S_p}\) denotes the surface delta distribution supported on
the port cross section. If the port coincides with a plane \(z=z_p\), it
may be written equivalently as \(\delta(z-z_p)\) restricted to the port
aperture.

\subsection{Floquet-port boundary condition}

For a periodic structure, we impose Bloch periodicity in the transverse
plane, $    \mathbf E(\boldsymbol\rho+\mathbf R,z)
    =
    e^{i\mathbf k_{\parallel}\cdot\mathbf R}
    \mathbf E(\boldsymbol\rho,z),$
where \(\mathbf R\) is a lattice vector and
\(\mathbf k_{\parallel}\) is the Bloch wave vector. For real
\(\mathbf k_{\parallel}\), the lateral Bloch boundary condition is
lossless and does not by itself generate a BA channel. Dissipation enters
through the open Floquet ports in the normal direction.

Let \(A_c\) be the unit-cell area and let \(\mathbf g\) denote a
reciprocal lattice vector. The tangential field on a top or bottom
Floquet port is expanded as
\begin{equation}
    \mathbf E_T^{(p)}(\boldsymbol\rho,\omega)
    =
    \sum_{\mathbf g,s}
    a_{p,\mathbf g s}(\omega)
    \mathbf e_{p,\mathbf g s}(\boldsymbol\rho,\omega),
\end{equation}
where
\begin{equation}
    \mathbf e_{p,\mathbf g s}(\boldsymbol\rho,\omega)
    =
    \frac{1}{\sqrt{A_c}}\,
    \hat{\mathbf e}^{(p)}_{\mathbf g s}
    e^{i(\mathbf k_{\parallel}+\mathbf g)\cdot\boldsymbol\rho}.
\end{equation}
Here \(s\in\{\mathrm{TE},\mathrm{TM}\}\) denotes the transverse
polarization. The normal wavenumber in the exterior medium is $    k_{z,\mathbf g}
    =
    \sqrt{
    k_b^2
    -
    |\mathbf k_{\parallel}+\mathbf g|^2
    },$
with the outgoing branch convention. A diffraction order is propagating
when \(k_{z,\mathbf g}\) is real and positive. Evanescent orders have
purely imaginary modal admittance and therefore contribute no positive
dissipation.

The Floquet-port condition has the same structure as the waveguide-port
condition,
\begin{equation}
    \hat{\mathbf n}\times\mathbf H
    =
    \sum_{\mathbf g,s}
    a_{p,\mathbf g s}
    Y_{p,\mathbf g s}
    \hat{\mathbf n}\times
    \left[
    \hat{\mathbf n}\times
    \mathbf e_{p,\mathbf g s}
    \right].
\end{equation}
Here \(Y_{p,\mathbf g s}\) denotes the physical modal admittance. For a
nonmagnetic homogeneous exterior medium with relative permittivity 
\(\epsilon_b\), $    Y_{\mathbf g,\mathrm{TE}}
    =
    \frac{k_{z,\mathbf g}}{\omega\mu_0},
    \qquad
    Y_{\mathbf g,\mathrm{TM}}
    =
    \frac{\omega\epsilon_0\epsilon_b}{k_{z,\mathbf g}} .$
More generally, if the exterior medium is magnetic, \(\mu_0\) and
\(\epsilon_0\epsilon_b\) should be replaced by the corresponding physical
permeability and permittivity. With the outgoing convention, passive
propagating orders have
\(\operatorname{Re}Y_{p,\mathbf g s}>0\).

The resulting Floquet BA channels are
\begin{equation}
    \mathbf C_{p,\mathbf g s}(\mathbf r,\omega)
    =
    \sqrt{
    \omega\mu_0
    \operatorname{Re}Y_{p,\mathbf g s}(\omega)
    }\,
    \mathbf e_{p,\mathbf g s}(\boldsymbol\rho,\omega)
    \delta_{S_p}(\mathbf r),
    \qquad
    (\mathbf g,s)\in\mathrm{prop}.
\end{equation}
These channels are the Rayleigh diffraction orders of the periodic open
system. Specular reflection and transmission correspond to
\(\mathbf g=0\), while higher-order diffraction channels correspond to
nonzero reciprocal lattice vectors satisfying the propagation condition.

\subsection{General passive open boundary conditions}

For an arbitrary passive open boundary condition, an analytic channel
basis may not be available. Once the open boundary condition is included
in the Maxwell boundary-value problem, the associated BA dissipative
operator can be numerically channelized.
Let \(\mathcal D_{\mathrm{BA}}(\omega)\) denote the BA dissipative
operator associated with the chosen open boundary condition. For a
passive boundary,
\begin{equation}
    \langle
    \mathbf E|
    \mathcal D_{\mathrm{BA}}(\omega)
    |\mathbf E
    \rangle
    \ge 0
\end{equation}
for every admissible field \(\mathbf E\). Therefore,
\(\mathcal D_{\mathrm{BA}}(\omega)\) admits the spectral decomposition
\begin{equation}
    \mathcal D_{\mathrm{BA}}(\omega)
    =
    \sum_n
    \lambda_n(\omega)
    \mathbf u_n(\omega)
    \mathbf u_n^\dagger(\omega),
    \qquad
    \lambda_n(\omega)\ge 0 .
    \label{eq:SM_general_BA_spectral}
\end{equation}
The corresponding numerical BA channel vectors are
\begin{equation}
    \mathbf C_n(\omega)
    =
    \sqrt{\lambda_n(\omega)}\,
    \mathbf u_n(\omega),
    \label{eq:SM_general_BA_channel}
\end{equation}
so that
\begin{equation}
    \mathcal D_{\mathrm{BA}}(\omega)
    =
    \sum_n
    \mathbf C_n(\omega)
    \mathbf C_n^\dagger(\omega).
\end{equation}
Eigenvalues below numerical tolerance are discarded, since they
correspond to nondissipative boundary degrees of freedom. Small negative
eigenvalues after Hermitian symmetrization indicate either numerical
roundoff or a non-passive boundary model.
In a finite-element implementation, this construction is applied to the
discrete BA dissipative matrix. If \(\mathbf B_{\rm obc}(\omega)\) is the
boundary contribution to the finite-element Maxwell matrix, then
\begin{equation}
    \mathbf D_{\mathrm{BA}}(\omega)
    =
    \frac{
    \mathbf B_{\rm obc}^{\dagger}(\omega)
    -
    \mathbf B_{\rm obc}(\omega)
    }{2i}.
\end{equation}
This is the discrete counterpart of
\(\mathcal D_{\mathrm{BA}}=-\operatorname{Im}\mathcal M_{\mathrm{BA}}\).
When the boundary degrees of freedom are represented in an orthonormal
basis, one may diagonalize
\begin{equation}
    \mathbf D_{\mathrm{BA}}(\omega)
    =
    \sum_n
    \lambda_n(\omega)
    \mathbf u_n(\omega)
    \mathbf u_n^\dagger(\omega),
\end{equation}
and define $    \mathbf c_n(\omega)
    =
    \sqrt{\lambda_n(\omega)}\,
    \mathbf u_n(\omega).$
If the boundary degrees of freedom are represented in a nonorthogonal
basis, the channelization should be understood in the
\(\mathbf M_{\partial}\)-weighted boundary Hilbert space. One solves the
generalized Hermitian eigenvalue problem
\begin{equation}
    \mathbf D_{\mathrm{BA}}\mathbf u_n
    =
    \lambda_n
    \mathbf M_{\partial}\mathbf u_n,
    \qquad
    \mathbf u_m^\dagger
    \mathbf M_{\partial}
    \mathbf u_n
    =
    \delta_{mn}.
\end{equation}
Then, for any boundary coefficient vector \(\mathbf x\),
\begin{equation}
    \mathbf x^\dagger
    \mathbf D_{\mathrm{BA}}
    \mathbf x
    =
    \sum_n
    \lambda_n
    \left|
    \mathbf u_n^\dagger
    \mathbf M_{\partial}
    \mathbf x
    \right|^2 .
\end{equation}
Equivalently, in a Euclidean coefficient representation,
\begin{equation}
    \mathbf D_{\mathrm{BA}}
    =
    \mathbf M_{\partial}
    \mathbf U
    \boldsymbol\Lambda
    \mathbf U^\dagger
    \mathbf M_{\partial},
\end{equation}
where the columns of \(\mathbf U\) are the generalized eigenvectors. The
corresponding channel functional acting on \(\mathbf x\) is 
\begin{equation}
    \mathbf c_n^\dagger \mathbf x
    =
    \sqrt{\lambda_n}\,
    \mathbf u_n^\dagger
    \mathbf M_{\partial}\mathbf x .
\end{equation}
This is the metric-consistent discrete analogue of the continuous Gram
factorization.
The numerical spectral channelization gives a valid bosonic reservoir
basis even when no closed-form radiation, port, Floquet, or
surface-admittance modes are known. The resulting channels are sufficient
for field quantization and for the Green-identity reconstruction $    \operatorname{Im}\mathbf G
    =
    \mathbf G
    \mathbf D
    \mathbf G^\dagger .$
However, numerical eigenchannels need not coincide with physically
labeled channels such as individual ports, diffraction orders, radiation
directions, or impedance-surface modes. They may be arbitrary unitary
mixtures within degenerate or nearly degenerate subspaces. 
To obtain physically resolved channels, one should first partition the
boundary degrees of freedom according to the desired measurement basis,
or factorize the dissipative form separately within the relevant port,
diffraction-order, radiation-direction, or impedance-surface subspaces. 
While the full channel sum is basis invariant, the interpretation of
individual channel contributions requires a physically chosen and
consistently indexed channel basis.

\section{Input-Output Theory}
\label{app:input_output_theory}

\subsection{Derivation of input-output relation}
\label{app:IO_derivation}

In this section, we provide a detailed input-output relation obtained by projecting the channel-resolved electric-field operator onto the monitored outgoing channels. The starting point is the channel decomposition of
the full dissipative operator,
\begin{equation}
    \mathcal D(\omega)
    =
    \mathcal D_{\rm ext}(\omega)
    +
    \mathcal D_{\rm int}(\omega),
    \qquad
    \mathcal D_{\eta}(\omega)
    =
    \mathcal C_{\eta}(\omega)
    \mathcal C_{\eta}^{\dagger}(\omega),
    \quad
    \eta\in\{{\rm ext},{\rm int}\}.
    \label{eq:SM_IO_D_split}
\end{equation}
The external channels are the monitored outgoing channels, such as
waveguide-port modes, Floquet diffraction orders, or selected
free-space radiation channels. The internal channels collect all
unmonitored dissipative degrees of freedom, including material
absorption, unobserved radiation, impedance-boundary loss, or any other
unmonitored passive channel.
The same external channel basis labels the incoming homogeneous
reservoir operators and the outgoing flux-normalized amplitudes. Thus
\(\hat{\mathbf a}_{\rm in}(\omega)\) denotes the incoming free
reservoir field associated with the monitored external channels, while
the outgoing amplitudes are obtained by projecting the total Maxwell
field onto the dual outgoing channel basis. The unmonitored internal
channels are described by independent bosonic operators
\(\hat{\mathbf f}_{\rm int}(\omega)\). We assume
\begin{equation}
\begin{gathered}
    \left[
    \hat{\mathbf a}_{\rm in}(\omega),
    \hat{\mathbf a}_{\rm in}^{\dagger}(\omega')
    \right]
    =
    \mathbf I\,\delta(\omega-\omega'),
    \qquad
    \left[
    \hat{\mathbf f}_{\rm int}(\omega),
    \hat{\mathbf f}_{\rm int}^{\dagger}(\omega')
    \right]
    =
    \mathbf I\,\delta(\omega-\omega'),
    \\
    \left[
    \hat{\mathbf a}_{\rm in}(\omega),
    \hat{\mathbf f}_{\rm int}^{\dagger}(\omega')
    \right]
    =
    0 .
\end{gathered}
\label{eq:SM_IO_channel_comm}
\end{equation}
The channel-resolved positive-frequency electric-field operator is
\begin{equation}
    \hat{\mathbf E}(\omega)
    =
    i\omega
    \sqrt{\frac{\hbar\mu_0}{\pi}}\,
    \mathcal G(\omega)
    \left[
        \mathcal C_{\rm ext}(\omega)
        \hat{\mathbf a}_{\rm in}(\omega)
        +
        \mathcal C_{\rm int}(\omega)
        \hat{\mathbf f}_{\rm int}(\omega)
    \right].
    \label{eq:SM_IO_field}
\end{equation}
This expression is field-centered: it gives the retarded Maxwell field
generated by all independent dissipative reservoir channels of the
open electromagnetic system.

Let \(\mathcal P_{\rm out}(\omega)\) denote the dual projection onto the
chosen flux-normalized outgoing external-channel basis. The outgoing
operators are defined by
\begin{equation}
    \hat{\mathbf a}_{\rm out}(\omega)
    =
    \mathcal P_{\rm out}(\omega)
    \hat{\mathbf E}(\omega).
    \label{eq:SM_IO_out_projection}
\end{equation}
For a waveguide port, \(\mathcal P_{\rm out}\) is realized as the
power-orthogonal modal projection onto the outgoing port mode. For
free-space or Floquet channels, it is the corresponding projection onto
outgoing radiation or diffraction modes. Its normalization is fixed by
the canonical photon-flux commutator,
\begin{equation}
    \left[
    \hat{\mathbf a}_{\rm out}(\omega),
    \hat{\mathbf a}_{\rm out}^{\dagger}(\omega')
    \right]
    =
    \mathbf I\,\delta(\omega-\omega')
    \label{eq:SM_IO_out_comm_norm}
\end{equation}
for the complete projected outgoing basis.
Substituting Eq.~\eqref{eq:SM_IO_field} into
Eq.~\eqref{eq:SM_IO_out_projection} gives the full-wave
frequency-domain input-output relation
\begin{equation}
    \hat{\mathbf a}_{\rm out}(\omega)
    =
    \mathbf S(\omega)\hat{\mathbf a}_{\rm in}(\omega)
    +
    \mathbf N(\omega)\hat{\mathbf f}_{\rm int}(\omega)
    \label{eq:SM_IO_frequency}
\end{equation}
with
\begin{align}
    \mathbf S(\omega)
    &=
    i\omega
    \sqrt{\frac{\hbar\mu_0}{\pi}}\,
    \mathcal P_{\rm out}(\omega)
    \mathcal G(\omega)
    \mathcal C_{\rm ext}(\omega),
    \label{eq:SM_S_def}
    \\
    \mathbf N(\omega)
    &=
    i\omega
    \sqrt{\frac{\hbar\mu_0}{\pi}}\,
    \mathcal P_{\rm out}(\omega)
    \mathcal G(\omega)
    \mathcal C_{\rm int}(\omega).
    \label{eq:SM_N_def}
\end{align}
The definition of \(\mathbf S(\omega)\) above assumes a channel convention in which the outgoing amplitude is obtained entirely from the projected total Green response. 
In channel conventions with an explicit prompt contribution, the scattering kernel should be understood as
\begin{equation}
    \mathbf S(\omega)
    =
    \mathbf S_{\rm dir}(\omega)
    +
    i\omega\sqrt{\frac{\hbar\mu_0}{\pi}}\,
    \mathcal P_{\rm out}(\omega)
    \mathcal G(\omega)
    \mathcal C_{\rm ext}(\omega),
    \label{eq:SM_S_with_direct}
\end{equation}
where \(\mathbf S_{\rm dir}\) is fixed by the normalization and phase convention of the chosen incoming/outgoing channel basis. 
For the port convention used in the FEM calculations, this direct term corresponds to the \(-\delta_{\alpha\mu}\) contribution in the projected port scattering formula. 
In the following, \(\mathbf S(\omega)\) denotes the total scattering kernel, including such direct terms when present.
Here \(\mathbf S(\omega)\) is the full electromagnetic map from incoming monitored external channels to outgoing monitored external channels, while \(\mathbf N(\omega)\) maps the unmonitored dissipative reservoir fluctuations to the monitored output channels.
After the usual analytic extension of the positive-frequency kernels,
or equivalently in a narrowband rotating-frame representation,
Eq.~\eqref{eq:SM_IO_frequency} may be written in the time domain as the
causal convolution
\begin{equation}
    \hat{\mathbf a}_{\rm out}(t)
    =
    \int_{-\infty}^{t}dt'\,
    \mathbf S(t-t')\hat{\mathbf a}_{\rm in}(t')
    +
    \int_{-\infty}^{t}dt'\,
    \mathbf N(t-t')\hat{\mathbf f}_{\rm int}(t'),
    \label{eq:SM_IO_time}
\end{equation}
where $    \mathbf S(t)
    =
    \int\frac{d\omega}{2\pi}\,
    \mathbf S(\omega)e^{-i\omega t},
\,    \mathbf N(t)
    =
    \int\frac{d\omega}{2\pi}\,
    \mathbf N(\omega)e^{-i\omega t}.$
The upper integration limit follows from the retarded nature of
\(\mathcal G(\omega)\). Unless \(\mathbf S(t)\) and \(\mathbf N(t)\)
collapse to local delta-function kernels, the output field depends on
the temporal history of both the incoming external field and the
unmonitored reservoir fluctuations. This is the full-wave origin of
non-Markovian input-output dynamics.

\subsection{Commutator preservation}
\label{app:IO_commutator}

The output commutator is preserved by the same Green identity that
ensures the field commutator. The electric-field operator satisfies
\begin{equation}
    \left[
    \hat{\mathbf E}(\omega),
    \hat{\mathbf E}^{\dagger}(\omega')
    \right]
    =
    \frac{\hbar\mu_0\omega^2}{\pi}
    \operatorname{Im}\mathcal G(\omega)\,
    \delta(\omega-\omega').
    \label{eq:SM_IO_E_comm}
\end{equation}
Using the channel decomposition
\(\mathcal D=\mathcal C_{\rm ext}\mathcal C_{\rm ext}^\dagger+
\mathcal C_{\rm int}\mathcal C_{\rm int}^\dagger\), the Green identity
becomes
\begin{equation}
    \operatorname{Im}\mathcal G
    =
    \mathcal G
    \mathcal C_{\rm ext}
    \mathcal C_{\rm ext}^{\dagger}
    \mathcal G^\dagger
    +
    \mathcal G
    \mathcal C_{\rm int}
    \mathcal C_{\rm int}^{\dagger}
    \mathcal G^\dagger .
    \label{eq:SM_IO_channel_Green_identity}
\end{equation}
Projecting Eq.~\eqref{eq:SM_IO_E_comm} onto the flux-normalized
outgoing basis gives
\begin{equation}
    \left[
    \hat{\mathbf a}_{\rm out}(\omega),
    \hat{\mathbf a}_{\rm out}^{\dagger}(\omega')
    \right]
    =
    \frac{\hbar\mu_0\omega^2}{\pi}
    \mathcal P_{\rm out}
    \operatorname{Im}\mathcal G
    \mathcal P_{\rm out}^{\dagger}
    \delta(\omega-\omega').
    \label{eq:SM_IO_projected_comm}
\end{equation}
Substituting Eq.~\eqref{eq:SM_IO_channel_Green_identity} and using the
definitions of \(\mathbf S\) and \(\mathbf N\), we obtain
\begin{equation}
    \left[
    \hat{\mathbf a}_{\rm out}(\omega),
    \hat{\mathbf a}_{\rm out}^{\dagger}(\omega')
    \right]
     =
    \left[
        \mathbf S(\omega)\mathbf S^\dagger(\omega)
        +
        \mathbf N(\omega)\mathbf N^\dagger(\omega)
    \right]
    \delta(\omega-\omega').
    \label{eq:SM_IO_projected_comm_SN}
\end{equation}
Comparison with the canonical output commutator,
\[
    [
    \hat{\mathbf a}_{\rm out}(\omega),
    \hat{\mathbf a}_{\rm out}^{\dagger}(\omega')
    ]
    =
    \mathbf I\,\delta(\omega-\omega'),
\]
therefore yields
\begin{equation}
    \mathbf S(\omega)\mathbf S^{\dagger}(\omega)
    +
    \mathbf N(\omega)\mathbf N^{\dagger}(\omega)
    =
    \mathbf I .
    \label{eq:SM_IO_commutator_preservation}
\end{equation}
This is the frequency-domain commutator-preservation condition for the
full-wave non-Markovian input-output map.
If all dissipative channels are monitored external channels,
\(\mathcal C_{\rm int}=0\), then \(\mathbf S\mathbf S^\dagger=\mathbf I\).
When unmonitored dissipation is present, \(\mathbf S\) is generally
subunitary, and the missing commutator is supplied by the noise map
\(\mathbf N\hat{\mathbf f}_{\rm int}\). Thus the noise operators in the
lossy input-output relation are the unmonitored dissipative channels of
the Maxwell operator, not phenomenological additions.
In the time domain, Eq.~\eqref{eq:SM_IO_commutator_preservation} is
equivalent to
\begin{equation}
    \int d\tau\,
    \left[
        \mathbf S(t-\tau)
        \mathbf S^{\dagger}(t'-\tau)
        +
        \mathbf N(t-\tau)
        \mathbf N^{\dagger}(t'-\tau)
    \right]
    =
    \mathbf I\,\delta(t-t').
    \label{eq:SM_IO_time_comm_preservation}
\end{equation}
The memory kernels generated by the full Maxwell response therefore
preserve the canonical output commutator exactly.

\section{Connection to Quasinormal-Mode Quantization}

In this section, we show that the QNM field operator and its
symmetrization used in a QNM quantization scheme~\cite{Franke_PRL_QNMQuantization_2019,QNM_PRA_2024,QNM_PRA_2022,QNM_Arxiv_2025} arise from a finite-pole projection of the
channel-resolved Maxwell field.
The exact positive-frequency field operator is
\begin{equation}
    \hat{\mathbf E}(\mathbf r,\omega)
    =
    i\omega
    \sqrt{\frac{\hbar\mu_0}{\pi}}
    \int d^3r'\,
    \mathbf G(\mathbf r,\mathbf r';\omega)
    \cdot
    \sum_j
    \mathbf C_j(\mathbf r',\omega)
    \hat{\mathbf b}_j(\omega).
    \label{eq:SM_qnm_exact_field}
\end{equation}
The index $j$ labels the independent dissipative channel groups, such
as boundary-assisted radiation and medium-assisted absorption.
In a frequency window dominated by a finite set of QNMs, we approximate
the Green tensor by
\begin{equation}
    \mathbf G(\mathbf r,\mathbf r';\omega)
    \simeq
    \sum_{\mu}
    A_{\mu}(\omega)
    \tilde{\mathbf f}_{\mu}(\mathbf r)
    \tilde{\mathbf f}_{\mu}(\mathbf r'),
    \label{eq:SM_qnm_green_expansion}
\end{equation}
where $\tilde{\mathbf f}_{\mu}$ is the QNM profile and
$\tilde{\omega}_{\mu}=\omega_{\mu}-i\gamma_{\mu}$ is the complex QNM
frequency.  The coefficient $A_{\mu}(\omega)$ depends on the chosen
QNM normalization convention.  Substituting
Eq.~\eqref{eq:SM_qnm_green_expansion} into
Eq.~\eqref{eq:SM_qnm_exact_field} gives
\begin{equation}
    \hat{\mathbf E}^{(+)}(\mathbf r)
    \simeq
    \sum_{\mu}
    \tilde{\mathbf f}_{\mu}(\mathbf r)
    \tilde{\alpha}_{\mu},
    \label{eq:SM_qnm_noncanonical_field}
\end{equation}
where the noncanonical QNM amplitudes are collective projections of
the underlying dissipative reservoirs,
\begin{equation}
    \tilde{\alpha}_{\mu}
    =
    \int_0^\infty d\omega\,
    A_{\mu}(\omega)
    \sum_j
    \left[
        \mathcal C_j^\dagger(\omega)
        |\tilde{\mathbf f}_{\mu}\rangle
    \right]^{\dagger}
    \hat{\mathbf b}_j(\omega),
    \label{eq:SM_qnm_alpha_from_channels}
\end{equation}
up to the same normalization convention used in
Eq.~\eqref{eq:SM_qnm_green_expansion}.  Thus the QNM amplitudes are not
postulated independently; they are finite-pole projections of the
boundary-assisted and medium-assisted reservoir operators.
Using the canonical commutation relations of the underlying channel
operators, one obtains
\begin{equation}
    [
        \tilde{\alpha}_{\mu},
        \tilde{\alpha}_{\eta}^{\dagger}
    ]
    =
    S_{\mu\eta},
    \label{eq:SM_qnm_S_def}
\end{equation}
with
\begin{equation}
    S_{\mu\eta}
    =
    \int_0^\infty d\omega\,
    A_{\mu}(\omega)A_{\eta}^{*}(\omega)
    \left\langle
        \tilde{\mathbf f}_{\mu}
        \left|
        \mathcal D(\omega)
        \right|
        \tilde{\mathbf f}_{\eta}
    \right\rangle .
    \label{eq:SM_qnm_S_from_D}
\end{equation}
Therefore the QNM commutator matrix is the finite-QNM Gram projection
of the dissipative Maxwell operator.

For a resonator in free space with material absorption,
\begin{equation}
    \mathcal D(\omega)
    =
    \mathcal D_{\rm BA}(\omega)
    +
    \mathcal D_{\rm MA}(\omega).
\end{equation}
Consequently,
\begin{equation}
    S_{\mu\eta}
    =
    S_{\mu\eta}^{\rm BA}
    +
    S_{\mu\eta}^{\rm MA},
\end{equation}
where
\begin{align}
    S_{\mu\eta}^{\rm BA}
    &=
    \int_0^\infty d\omega\,
    A_{\mu}A_{\eta}^{*}
    \left\langle
        \tilde{\mathbf f}_{\mu}
        \left|
        \mathcal D_{\rm BA}
        \right|
        \tilde{\mathbf f}_{\eta}
    \right\rangle,
    \\
    S_{\mu\eta}^{\rm MA}
    &=
    \int_0^\infty d\omega\,
    A_{\mu}A_{\eta}^{*}
    \left\langle
        \tilde{\mathbf f}_{\mu}
        \left|
        \mathcal D_{\rm MA}
        \right|
        \tilde{\mathbf f}_{\eta}
    \right\rangle .
\end{align}
Thus the conventional radiative and nonradiative QNM overlaps are
identified as
\begin{equation}
    S_{\mu\eta}^{\rm rad}
    \equiv
    S_{\mu\eta}^{\rm BA},
    \qquad
    S_{\mu\eta}^{\rm nrad}
    \equiv
    S_{\mu\eta}^{\rm MA}.
\end{equation}
This identification shows that the usual radiative/nonradiative split
of QNM quantization is the free-space realization of the general
boundary-assisted/medium-assisted decomposition of the Maxwell
dissipation operator.

Since $S_{\mu\eta}$ is generally not the identity, the amplitudes
$\tilde{\alpha}_{\mu}$ do not define canonical bosonic modes.  We
therefore introduce the canonical QNM operators
\begin{equation}
    \hat a_{\mu}
    =
    \sum_{\nu}
    (S^{-1/2})_{\mu\nu}
    \tilde{\alpha}_{\nu},
    \qquad
    [
        \hat a_{\mu},
        \hat a_{\eta}^{\dagger}
    ]
    =
    \delta_{\mu\eta}.
\end{equation}
Equivalently,
\begin{equation}
    \tilde{\alpha}_{\mu}
    =
    \sum_{\nu}
    (S^{1/2})_{\mu\nu}
    \hat a_{\nu}.
\end{equation}
Substituting this relation into
Eq.~\eqref{eq:SM_qnm_noncanonical_field} gives the symmetrized QNM
field operator
\begin{equation}
    \hat{\mathbf E}^{(+)}(\mathbf r)
    \simeq
    \sum_{\nu}
    \tilde{\mathbf f}^{\,s}_{\nu}(\mathbf r)
    \hat a_{\nu},
    \label{eq:SM_qnm_sym_field_compact}
\end{equation}
where
\begin{equation}
    \tilde{\mathbf f}^{\,s}_{\nu}(\mathbf r)
    =
    \sum_{\mu}
    \tilde{\mathbf f}_{\mu}(\mathbf r)
    (S^{1/2})_{\mu\nu}.
    \label{eq:SM_qnm_sym_mode}
\end{equation}
Restoring the conventional QNM normalization gives the familiar form
\begin{equation}
    \hat{\mathbf E}(\mathbf r)
    =
    i
    \sqrt{\frac{\hbar}{2\epsilon_0}}
    \sum_{\nu}
    \sqrt{\omega_{\nu}}\,
    \tilde{\mathbf f}^{\,s}_{\nu}(\mathbf r)
    \hat a_{\nu}
    +
    {\rm H.a.}
    \label{eq:SM_qnm_field_operator}
\end{equation}
Thus the symmetrized QNM field operator is recovered as the finite-pole
projection of the exact dissipative-channel field operator.
The present derivation also clarifies the meaning of QNM
symmetrization.  The matrix $S_{\mu\eta}$ is not an additional
postulate of QNM quantization; it is the projection of
$\mathcal D=-\operatorname{Im}\mathcal M$ onto the QNM subspace.
Moreover, its radiative and nonradiative parts are the projections of
$\mathcal D_{\rm BA}$ and $\mathcal D_{\rm MA}$, respectively.  For
more general open systems, the same construction resolves the QNM
field operator into port, impedance, Floquet, radiation, and material
loss channels whenever the corresponding dissipative operators are
available.
\section{Numerical Simulation Details}
\label{app:simulation_details}
This section provides the detailed numerical setup and implementation for the simulation results presented in the main text. To validate the channel-resolved quantization framework, we perform a finite-element method (FEM) simulation of a one-dimensional open waveguide. 

\subsection{Numerical example 1: Impedance and outgoing boundary conditions}
\begin{figure}[htbp]
    \centering
    \includegraphics[width=0.4\linewidth]{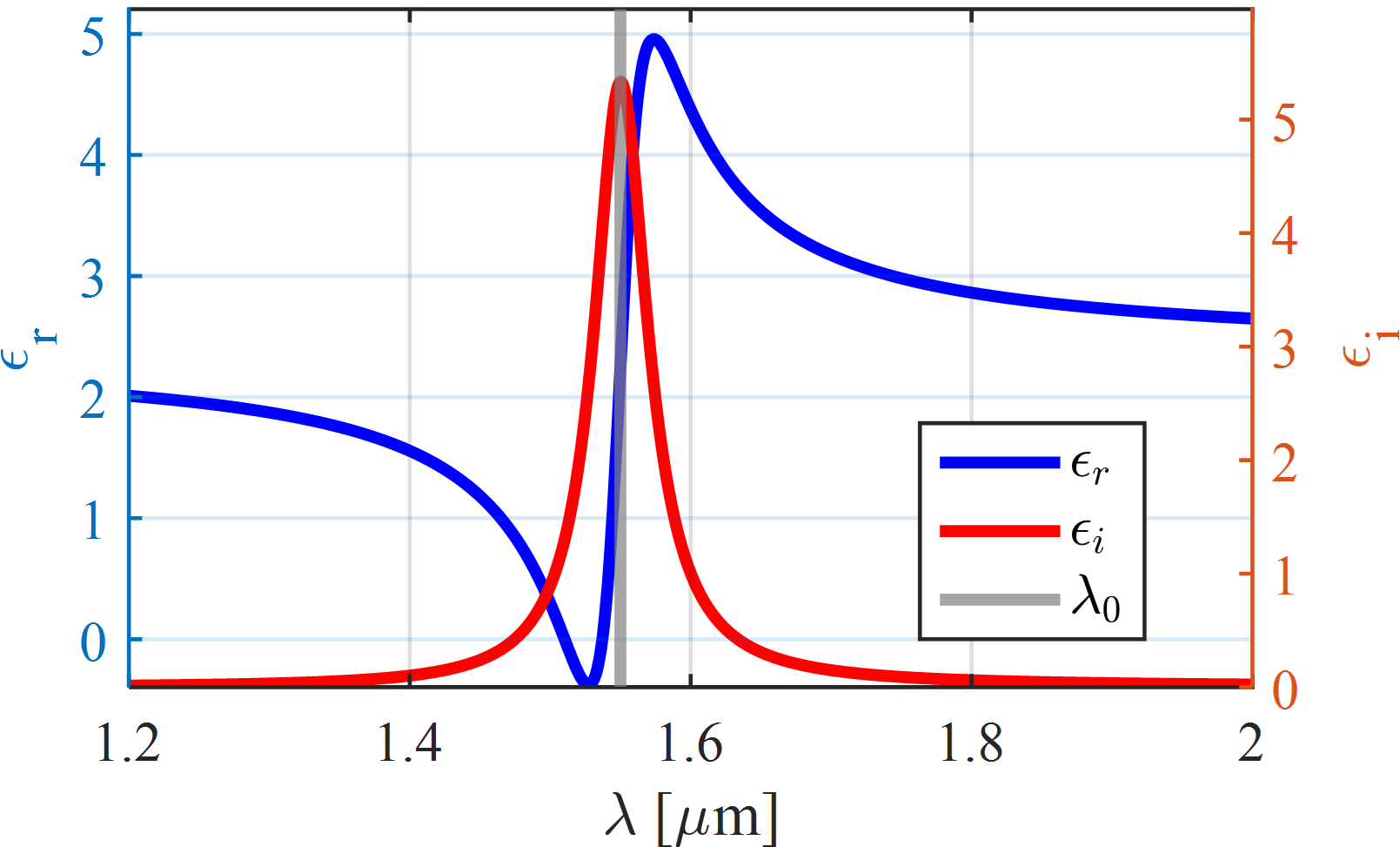}
    \caption{
    Lorentz permittivity used for the lossy slab in Numerical example 1.
    The real and imaginary parts of $\epsilon_{\rm slab}(\omega)$ are plotted as functions of wavelength.
    The vertical dashed line marks the resonance wavelength $\lambda_0=1.55\,\mu{\rm m}$.
    }
    \label{fig:lorentz_eps_SM}
\end{figure}

\begin{figure}[htbp]
    \centering
    \begin{minipage}[b]{0.4\textwidth}
        \centering
        \includegraphics[width=\textwidth]{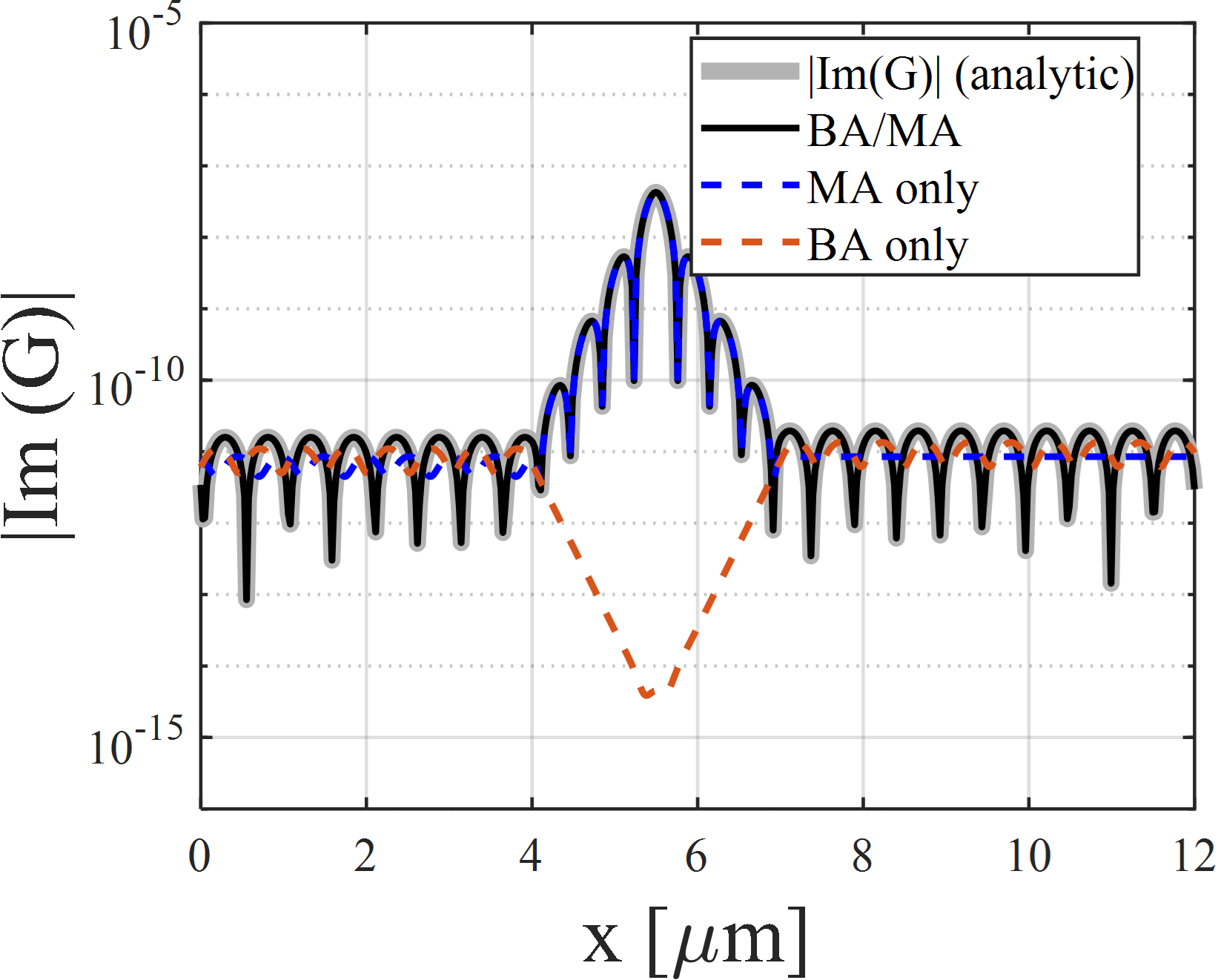}
        \par\smallskip
        \textbf{(a)}
        \label{fig:ImG_inside}
    \end{minipage}
    \begin{minipage}[b]{0.4\textwidth}
        \centering
        \includegraphics[width=\textwidth]{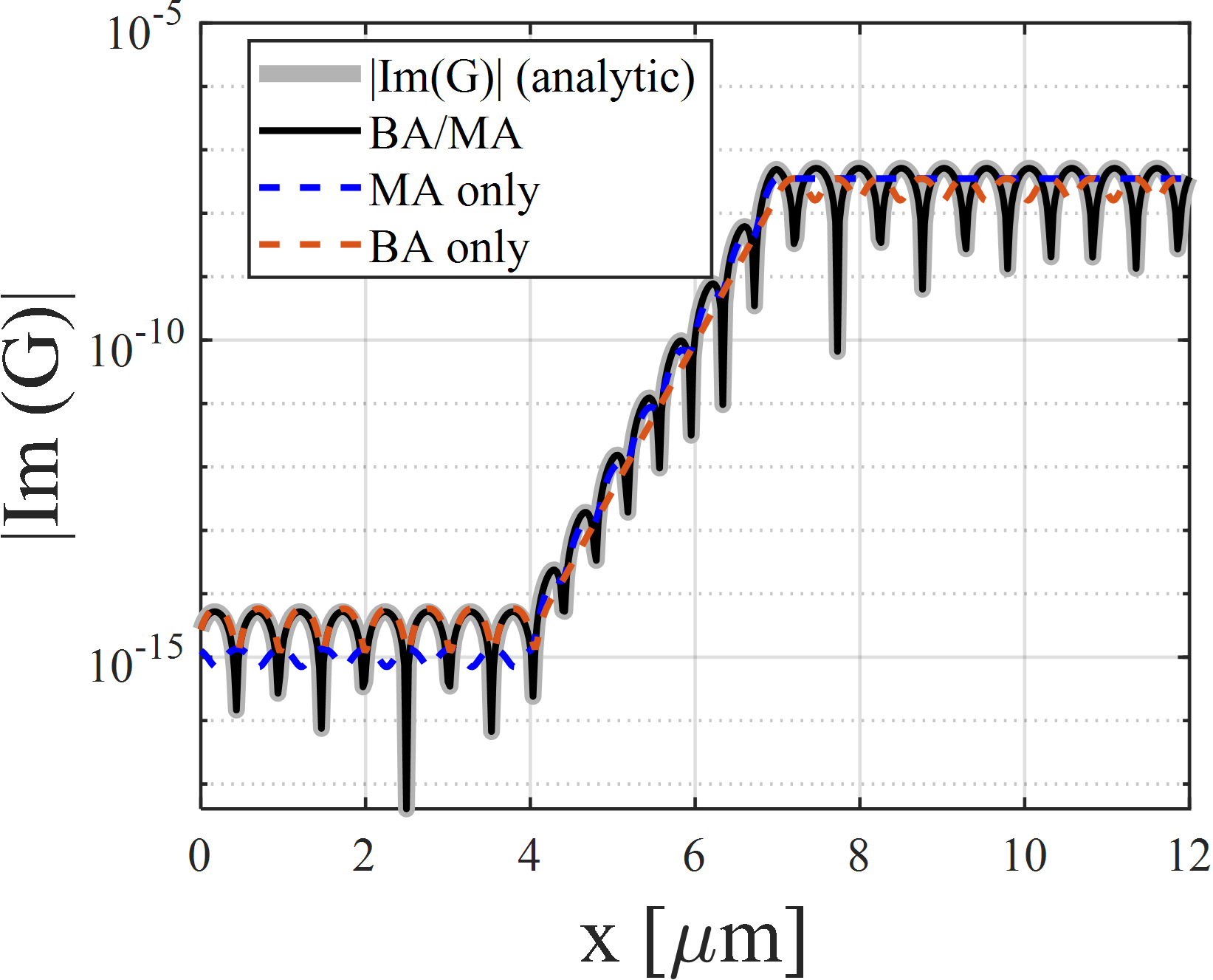}
        \par\smallskip
        \textbf{(b)}
        \label{fig:ImG_outside}
    \end{minipage}

    \caption{Spatial distribution of $|\operatorname{Im}(G)|$ for an emitter (a) inside ($x_0 = 5.5\,\mu\text{m}$) and (b) outside ($x_0 = 8.5\,\mu\text{m}$) the lossy slab. The full BA/MA reconstruction (black solid) exactly reproduces the analytic reference (grey).}
    \label{fig:ImG_combined}
\end{figure}

 A lossy dispersive dielectric slab is embedded in the region $x \in [4.0, 7.0]\,\mu\text{m}$. The permittivity of the slab is modeled using a Lorentz susceptibility, $\epsilon_{\text{slab}}(\omega) = \epsilon_{\text{bg}} + \frac{\omega_p^2}{\omega_0^2 - \omega^2 - i\gamma\omega},
$
where $\omega_0$ is the resonance frequency corresponding to a wavelength of $\lambda_0 = 1.55\,\mu\text{m}$. The plasma frequency is set to $\omega_p = 0.40\omega_0$, and the damping rate is $\gamma = 0.03\omega_0$.
The waveguide is terminated by distinct open boundary conditions at both ends. At the right boundary ($x = L$), an exact outgoing Dirichlet-to-Neumann (DtN) boundary condition is imposed, simulating radiation into an infinite uniform background. At the left boundary ($x = 0$), a passive impedance boundary condition (IBC) is applied with a normalized load $z_L = Z_L / Z_c = 1.30 + 0.70i$, where $Z_c$ is the characteristic impedance of the background medium. Figure~\ref{fig:lorentz_eps_SM} shows the corresponding real and imaginary parts of the slab permittivity.
To verify the completeness of the channel factorization, we calculate the imaginary part of the Green's function, which dictates the local density of states and the fluctuation spectrum. For a point source located at $x_0$, the exact numerical reconstruction is given by the quadratic form $\operatorname{Im} \mathbf{G} = \mathbf{z}^\dagger \mathbf{D} \mathbf{z}$, where $\mathbf{z}$ is the adjoint field corresponding to the column of the inverse Maxwell matrix $\mathbf{A}^{-1}$.
Figure~\ref{fig:ImG_combined} shows the spatial distribution of $|\operatorname{Im}(\mathbf{G})|$ evaluated at the resonance frequency for an emitter placed (a) inside the lossy slab ($x_0 = 5.5\,\mu\text{m}$) and (b) outside the slab ($x_0 = 8.5\,\mu\text{m}$). The full BA/MA channel sum perfectly matches the exact semianalytic Green's function across the entire domain. Notably, performing partial reconstructions highlights the necessity of the unified framework: for the emitter inside the slab [Fig.~\ref{fig:ImG_combined}(a)], MA channels dominate the local dissipation, whereas for the emitter outside [Fig.~\ref{fig:ImG_combined}(b)], the BA channels (radiation and impedance load) are strictly required to correctly reconstruct the long-range electromagnetic fluctuations.

\subsection{Numerical example 2: photonic integrated circuit configuration}

\begin{figure*}[t]
    \centering

    \begin{minipage}[t]{0.5\textwidth}
        \centering
        \includegraphics[width=\linewidth]{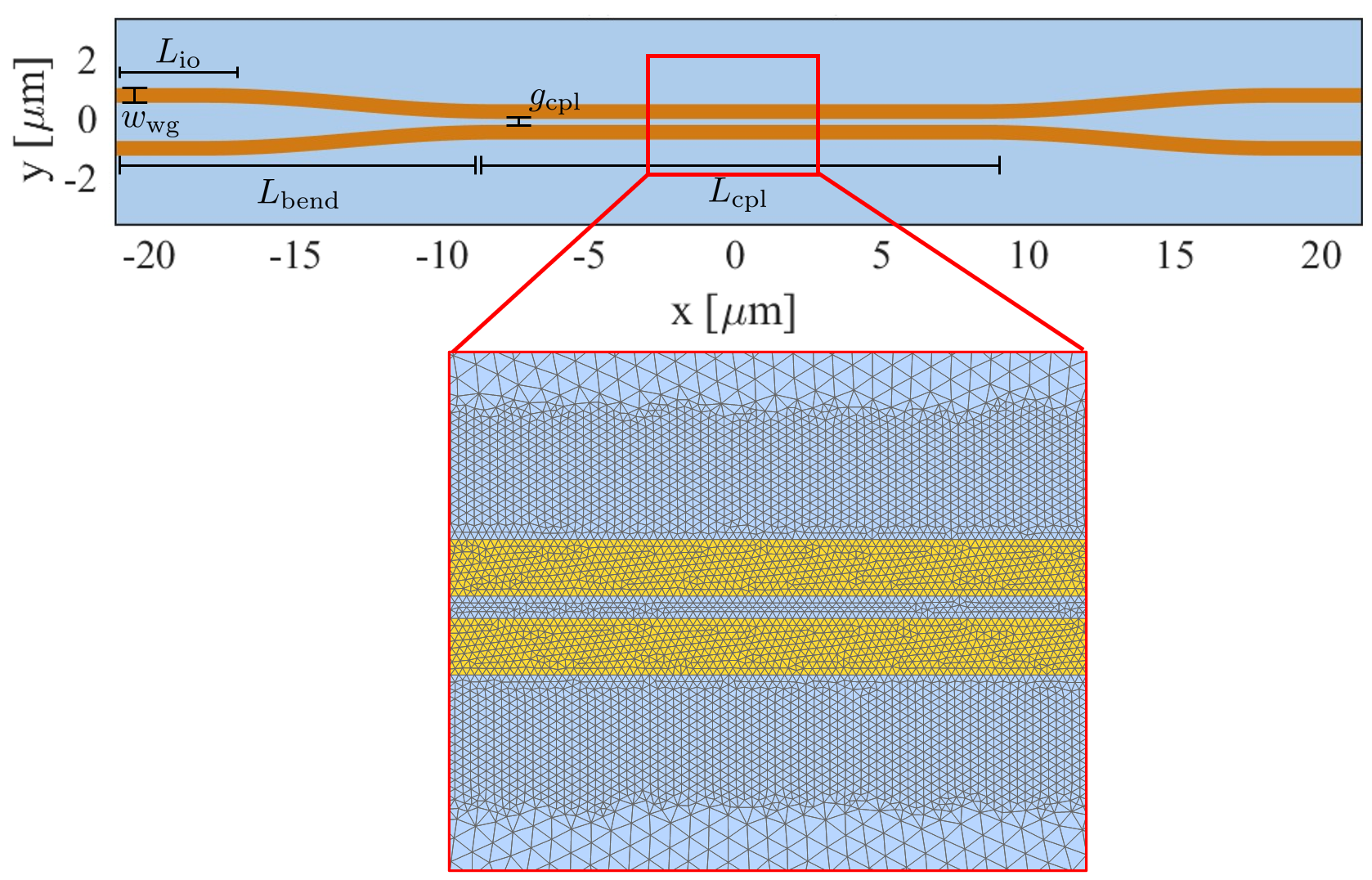}
        \vspace{-1.0ex}

        \textbf{(a)}
    \end{minipage}%
    \hfill
    \begin{minipage}[t]{0.45\textwidth}
        \centering
        \includegraphics[width=\linewidth]{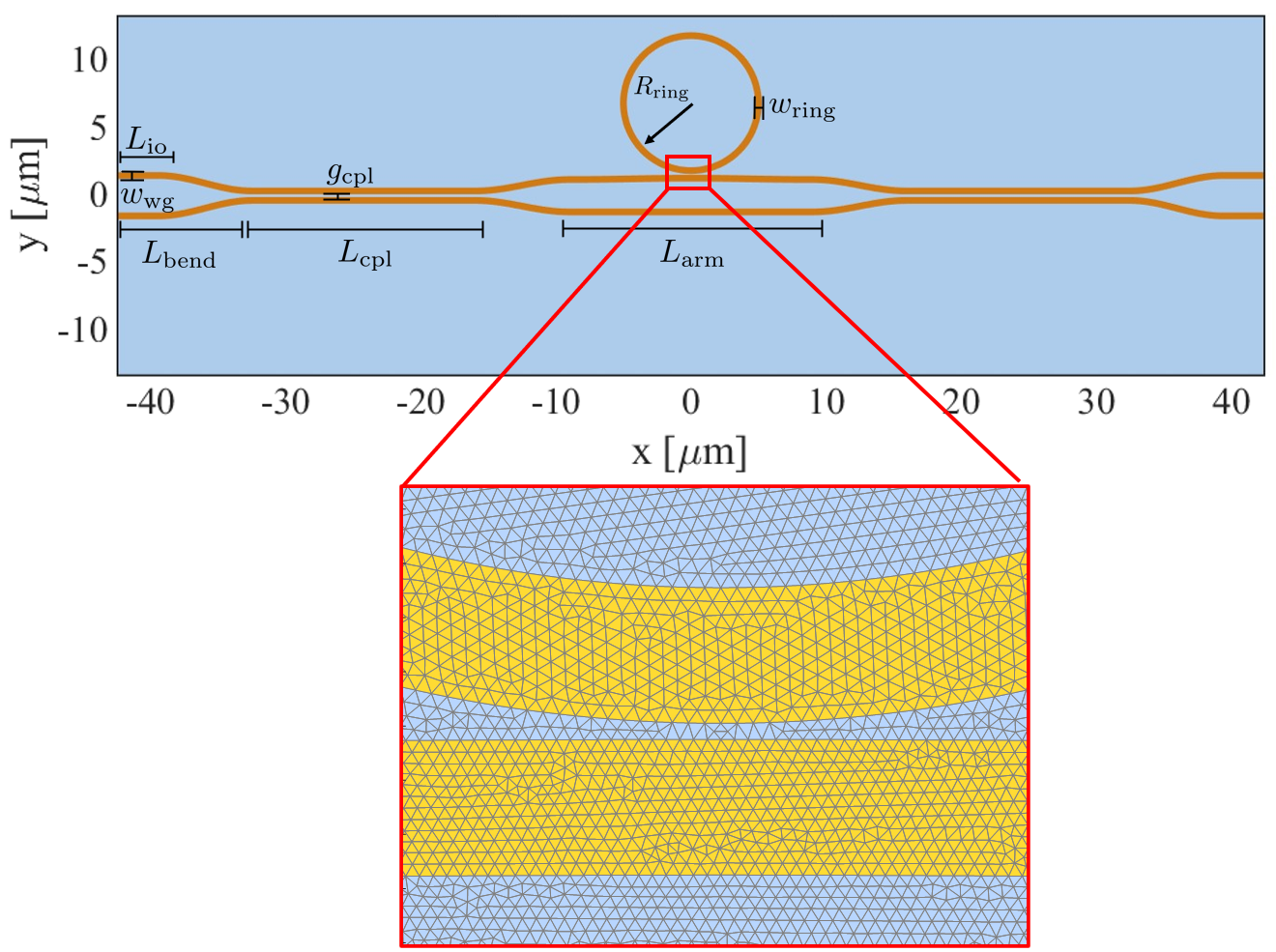}
        \vspace{-1.0ex}

        \textbf{(b)}
    \end{minipage}

    \caption{
    Finite-element meshes used for the PIC examples.
    (a) Broadband 50/50 splitter.
    (b) Ring-loaded interferometric splitter.
    Yellow regions denote high-index waveguide cores and blue regions denote the cladding/background. Red boxes indicate the zoomed regions where the triangular mesh is shown explicitly.
    }
    \label{fig:SM_pic_mesh}
\end{figure*}

The calculations are performed with a two-dimensional scalar effective-index finite-element model of the planar photonic integrated circuit geometries. 
We consider two devices: a broadband directional coupler, used as a spectrally smooth Markovian-limit control, and a ring-loaded interferometric splitter, used to generate a structured port response. 
The geometrical, material, and boundary parameters are summarized in Table~\ref{tab:pic_parameters}, and the corresponding finite-element meshes are shown in Fig.~\ref{fig:SM_pic_mesh}.
Material absorption is included directly in the Maxwell operator through the complex effective permittivity, $    \epsilon({\bf r})
    =
    \epsilon'({\bf r})+i\epsilon''({\bf r}).$
The imaginary part of the permittivity defines the medium-assisted contribution $D_{\rm MA}$, while the waveguide-port boundary loading defines the boundary-assisted contribution $D_{\rm BA}$. 
Thus the PIC loss is not introduced as a post-processing attenuation factor; it is part of the same lossy finite-element Green response used to compute the scattering matrix.

Both PIC configurations use the same local slab-port cross section, with $w_{\rm wg}=0.50\,\mu{\rm m}$, $n_{\rm core}=2.58$, and $n_{\rm clad}=1.444$. 
The local port dispersion supports two guided modes throughout the operating wavelength band, while higher-order candidates are below cutoff. 
Accordingly, both the broadband coupler and the ring-loaded splitter are represented using the same two-channel port basis at each physical port. 
The second guided mode is included as a boundary-assisted channel of the open Maxwell problem; it is not populated in the input state unless explicitly excited.
\begin{table}[b]
\centering
\caption{
Geometry, material, and boundary parameters used for the PIC examples.
}
\small
\begin{tabular}{lcc}
\hline
Parameter & Broadband 50/50 splitter & Ring-loaded splitter \\
\hline
Waveguide width $w_{\rm wg}$ 
& $0.50\,\mu{\rm m}$ 
& $0.50\,\mu{\rm m}$ \\
Coupler gap $g_{\rm cpl}$ 
& $0.20\,\mu{\rm m}$ 
& $0.20\,\mu{\rm m}$ \\
Coupler length $L_{\rm cpl}$ 
& $16.5\,\mu{\rm m}$ 
& $16.5\,\mu{\rm m}$ \\
Input/output straight length $L_{\rm io}$ 
& $3.0\,\mu{\rm m}$ 
& $3.0\,\mu{\rm m}$ \\
Bend length $L_{\rm bend}$ 
& $10.0\,\mu{\rm m}$ 
& $7.0\,\mu{\rm m}$ \\
Port center separation 
& $1.80\,\mu{\rm m}$ 
& $3.00\,\mu{\rm m}$ \\
Interferometer arm separation 
& -- 
& $2.40\,\mu{\rm m}$ \\
Arm length $L_{\rm arm}$ 
& -- 
& $18.0\,\mu{\rm m}$ \\
Ring width $w_{\rm ring}$ 
& -- 
& $0.50\,\mu{\rm m}$ \\
Ring radius $R_{\rm ring}$ 
& -- 
& $5.00\,\mu{\rm m}$ \\
Ring-arm gap 
& -- 
& $0.16\,\mu{\rm m}$ \\
Upper-arm phase bump 
& -- 
& $0.100\,\mu{\rm m}$ \\
Center wavelength $\lambda_c$ 
& $1550\,{\rm nm}$ 
& $1557.731698\,{\rm nm}$ \\
$n_{\rm core}$, $n_{\rm clad}$ 
& $2.58$, $1.444$ 
& $2.58$, $1.444$ \\
$\epsilon''_{\rm core}$, $\epsilon''_{\rm clad}$ 
& $2.0\times10^{-4}$, $2.0\times10^{-5}$
& $2.0\times10^{-4}$, $2.0\times10^{-5}$ \\
Retained guided modes per physical port
& 2
& 2 \\
\hline
\end{tabular}
\label{tab:pic_parameters}
\end{table}
For each guided port mode $\alpha=(p,m)$, where $p$ labels the physical port and $m$ labels the transverse guided mode, the local transverse profile is projected onto the finite-element boundary to define a normalized dual vector $w_\alpha(\omega)$. 
The port contribution to the dissipative Maxwell operator is
\begin{align}
    D_{\rm port}(\omega)
    &=
    \sum_{\alpha}
    \beta_\alpha(\omega)
    w_\alpha(\omega)w_\alpha^\dagger(\omega),
    \label{eq:SM_Dport}
\end{align}
where $\beta_\alpha$ is the propagation constant of the guided port mode. 
The projected scattering matrix is computed from the full lossy finite-element Green response as 
\begin{align}
    S_{\alpha\mu}(\omega)
    &=
    -2i\,\beta_\mu(\omega)\,
    w_\alpha^\dagger(\omega)
    {\bf G}_{\rm FEM}(\omega)
    w_\mu(\omega)
    -
    \delta_{\alpha\mu}.
    \label{eq:SM_port_S}
\end{align}
Here ${\bf G}_{\rm FEM}(\omega)$ includes the actual two-dimensional device geometry, the imaginary permittivity, and the modal port loading. 
Equivalently, in the implementation one solves the sparse finite-element system for the port-source right-hand sides and evaluates $    S(\omega)
    =
    W_{\rm out}^\dagger(\omega)
    {\bf G}_{\rm FEM}(\omega)
    B_{\rm in}(\omega)
    -
    I .
    \label{eq:SM_S_matrix}$
The full finite-bandwidth calculation uses the frequency-dependent scattering matrix itself, while the Markovian reference freezes the same lossy device at the center frequency, $    S_{\rm NM}(\omega)
    =
    S(\omega),\,
    S_{\rm M}(\omega)
    =
    S(\omega_c).$
The two-photon input state is prepared only in the fundamental guided modes of the two left physical ports,
\begin{align}
    |\psi_{\rm in}\rangle
    &=
    \left[
    \int d\omega\,
    \xi(\omega)\,
    \hat a_{L_1,m=1}^\dagger(\omega)
    \right]
    \left[
    \int d\omega\,
    \xi(\omega)\,
    \hat a_{L_2,m=1}^\dagger(\omega)
    \right]
    |0\rangle .
    \label{eq:SM_input_state}
\end{align}
The retained second guided modes in the input ports are therefore vacuum channels. 
For the output observable we use physical-port bucket detection: all retained guided channels leaving the upper right physical port define the output set $\mathcal A$, and all retained guided channels leaving the lower right physical port define the output set $\mathcal B$. 
Thus the calculation describes a single-mode two-photon input propagating through a multi-channel lossy port environment, followed by bucket detection over the retained guided output channels.
The single-photon spectrum is a normalized Gaussian,
\begin{align}
    \xi(\omega)
    &=
    \mathcal N
    \exp\left[
    -\frac{\sigma_t^2(\omega-\omega_c)^2}{2}
    \right],
    &
    \int d\omega\,|\xi(\omega)|^2
    &=
    1 .
    \label{eq:SM_photon_spectrum}
\end{align}
For output sets $\mathcal A$ and $\mathcal B$, the distinguishable coincidence probability is
\begin{align}
    P_{\rm dist}
    &=
    I_{\mathcal A1}I_{\mathcal B2}
    +
    I_{\mathcal B1}I_{\mathcal A2},
    \\
    I_{\mathcal A j}
    &=
    \sum_{a\in\mathcal A}
    \int d\omega\,
    |\xi(\omega)|^2
    |S_{aj}(\omega)|^2,
    \\
    I_{\mathcal B j}
    &=
    \sum_{b\in\mathcal B}
    \int d\omega\,
    |\xi(\omega)|^2
    |S_{bj}(\omega)|^2 .
    \label{eq:SM_Pdist}
\end{align}
For indistinguishable photons with relative delay $\tau$, the coherent two-photon overlap gives
\begin{align}
    P_{\rm coin}(\tau)
    &=
    P_{\rm dist}
    +
    2\,{\rm Re}
    \left[
    \Gamma_{\mathcal A}(\tau)
    \Gamma_{\mathcal B}(\tau)
    \right],
    \\
    \Gamma_{\mathcal A}(\tau)
    &=
    \sum_{a\in\mathcal A}
    \int d\omega\,
    |\xi(\omega)|^2
    e^{i\omega\tau}
    S_{a1}(\omega)
    S^*_{a2}(\omega),
    \\
    \Gamma_{\mathcal B}(\tau)
    &=
    \sum_{b\in\mathcal B}
    \int d\omega\,
    |\xi(\omega)|^2
    e^{-i\omega\tau}
    S_{b2}(\omega)
    S^*_{b1}(\omega).
    \label{eq:SM_Pcoin}
\end{align}
The normalized HOM correlation is
\begin{align}
    g^{(2)}(\tau)
    &=
    \frac{P_{\rm coin}(\tau)}{P_{\rm dist}} .
    \label{eq:SM_g2}
\end{align}
Material-loss reservoirs are taken to be in vacuum, so they do not add normally ordered detector counts; their effect is already contained in the lossy, subunitary scattering matrix $S(\omega)$.

For spectral diagnostics we evaluate the HOM interference kernel
\begin{align}
    K(\omega)
    &=
    \sum_{a\in\mathcal A}
    \sum_{b\in\mathcal B}
    \left[
    S_{a1}(\omega)S_{b2}(\omega)
    +
    S_{b1}(\omega)S_{a2}(\omega)
    \right].
    \label{eq:SM_K}
\end{align}
For single-mode output ports, this reduces to
\begin{align}
    K(\omega)
    &=
    S_{31}(\omega)S_{42}(\omega)
    +
    S_{41}(\omega)S_{32}(\omega).
    \label{eq:SM_K_single_mode}
\end{align}
The broadband coupler has a nearly flat $K(\omega)$ over the photon spectrum, and therefore its finite-bandwidth response coincides with the flat-$S(\omega_c)$ reference. 
In contrast, the ring-loaded splitter is tuned close to monochromatic HOM cancellation at $\omega_c$, but the ring produces strong spectral variation of $S(\omega)$ and $K(\omega)$ across the photon bandwidth. 
This finite-bandwidth sampling of the structured port Green response produces the reduced and shifted HOM dip shown in the main text.
\begin{figure*}
    \centering
    \begin{minipage}[t]{0.4\textwidth}
        \centering
        \includegraphics[width=\linewidth]{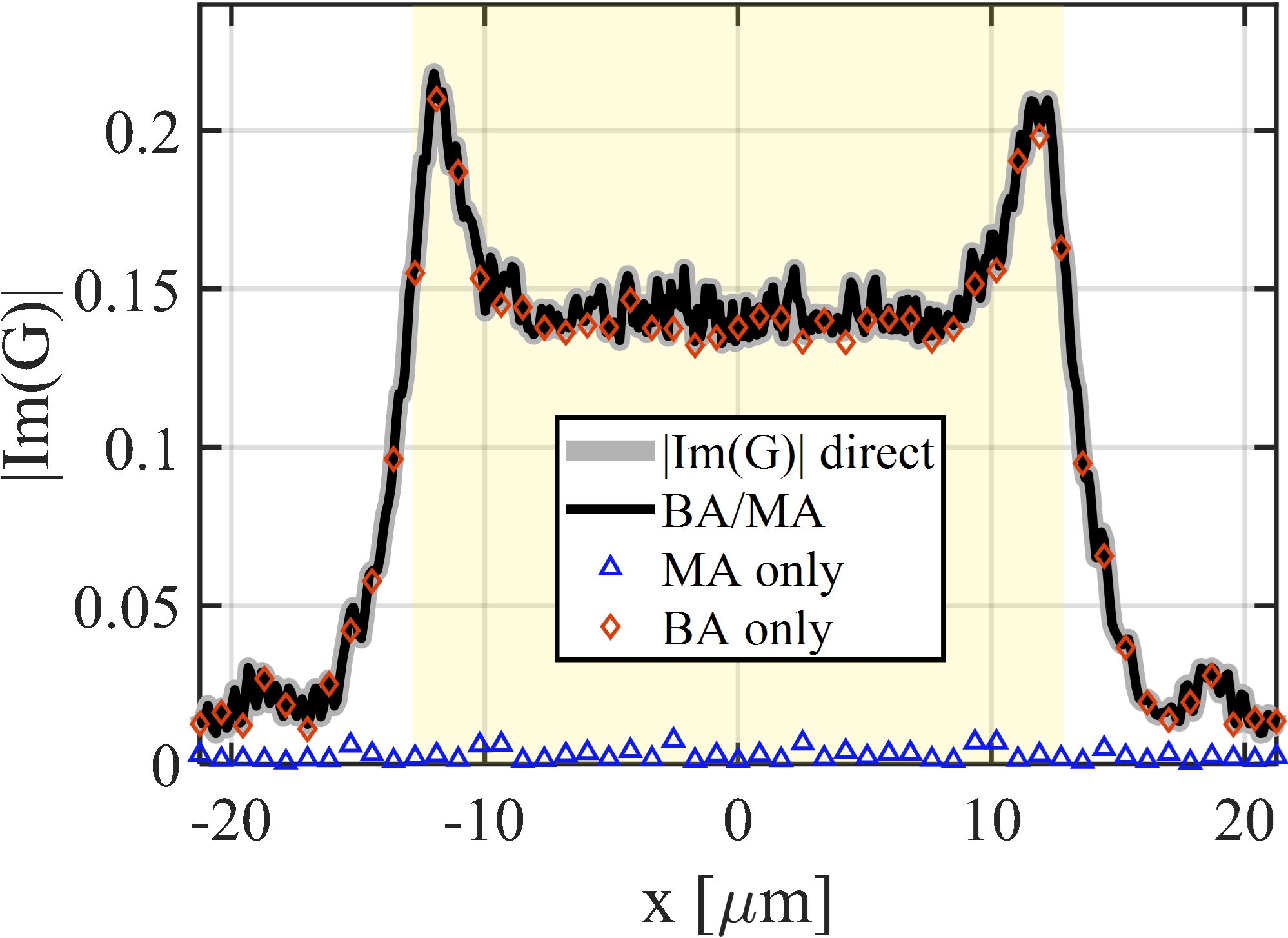}
        \vspace{-1.0ex}

        \textbf{(a)}
    \end{minipage}
    \begin{minipage}[t]{0.4\textwidth}
        \centering
        \includegraphics[width=\linewidth]{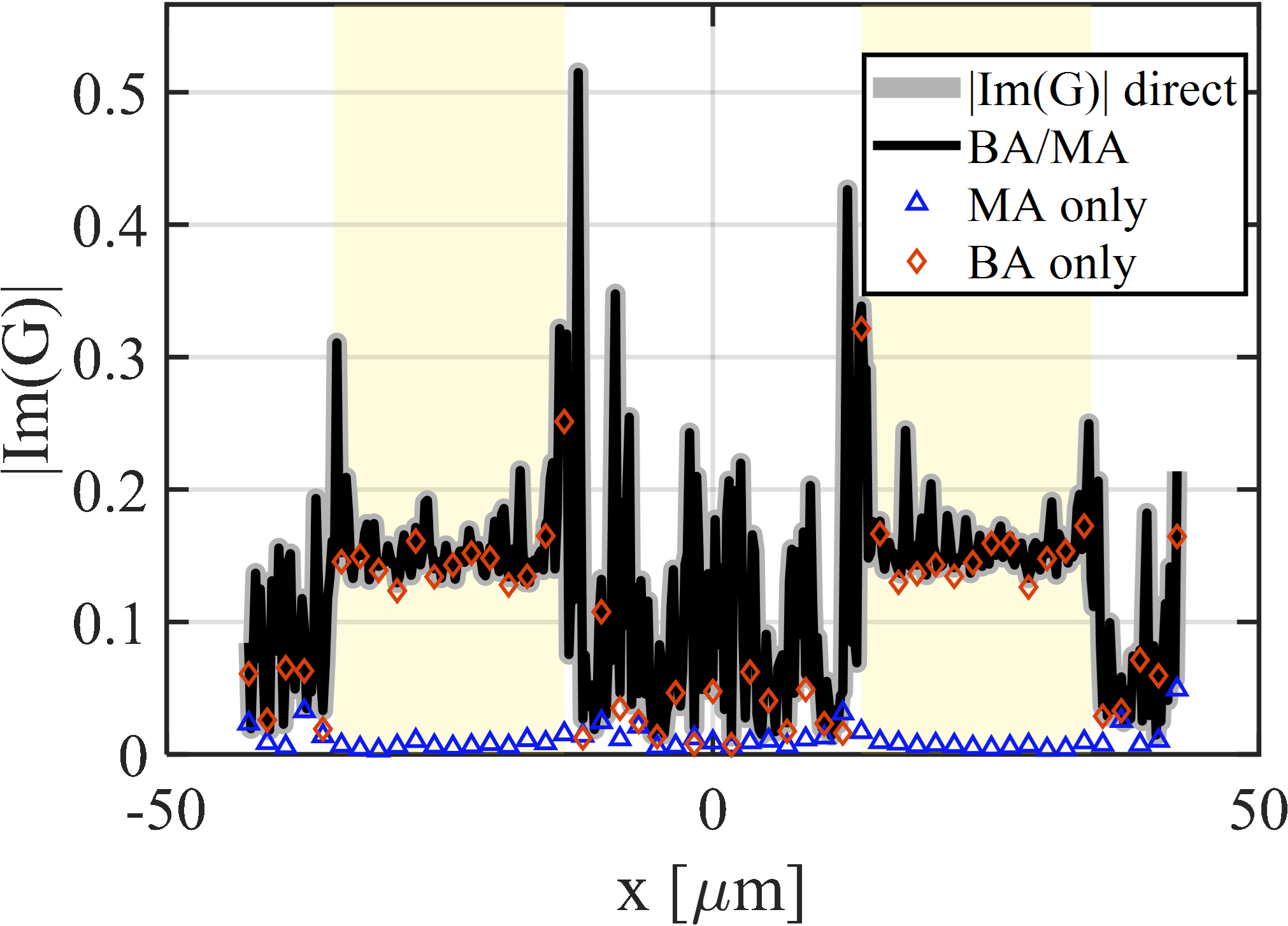}
        \vspace{-1.0ex}

        \textbf{(b)}
    \end{minipage}

    \begin{minipage}[t]{0.4\textwidth}
        \centering
        \includegraphics[width=\linewidth]{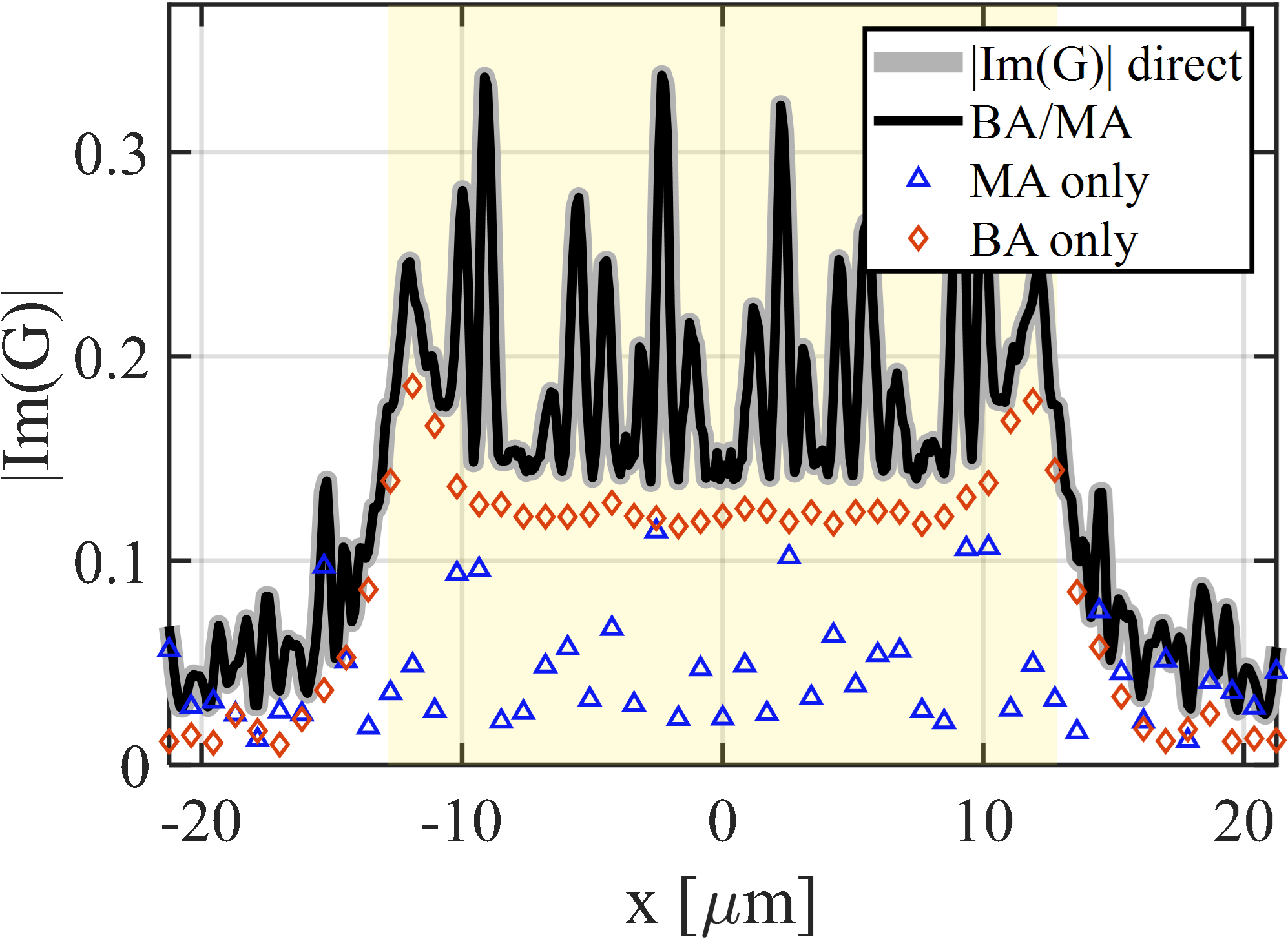}
        \vspace{-1.0ex}

        \textbf{(c)}
    \end{minipage}
    \begin{minipage}[t]{0.4\textwidth}
        \centering
        \includegraphics[width=\linewidth]{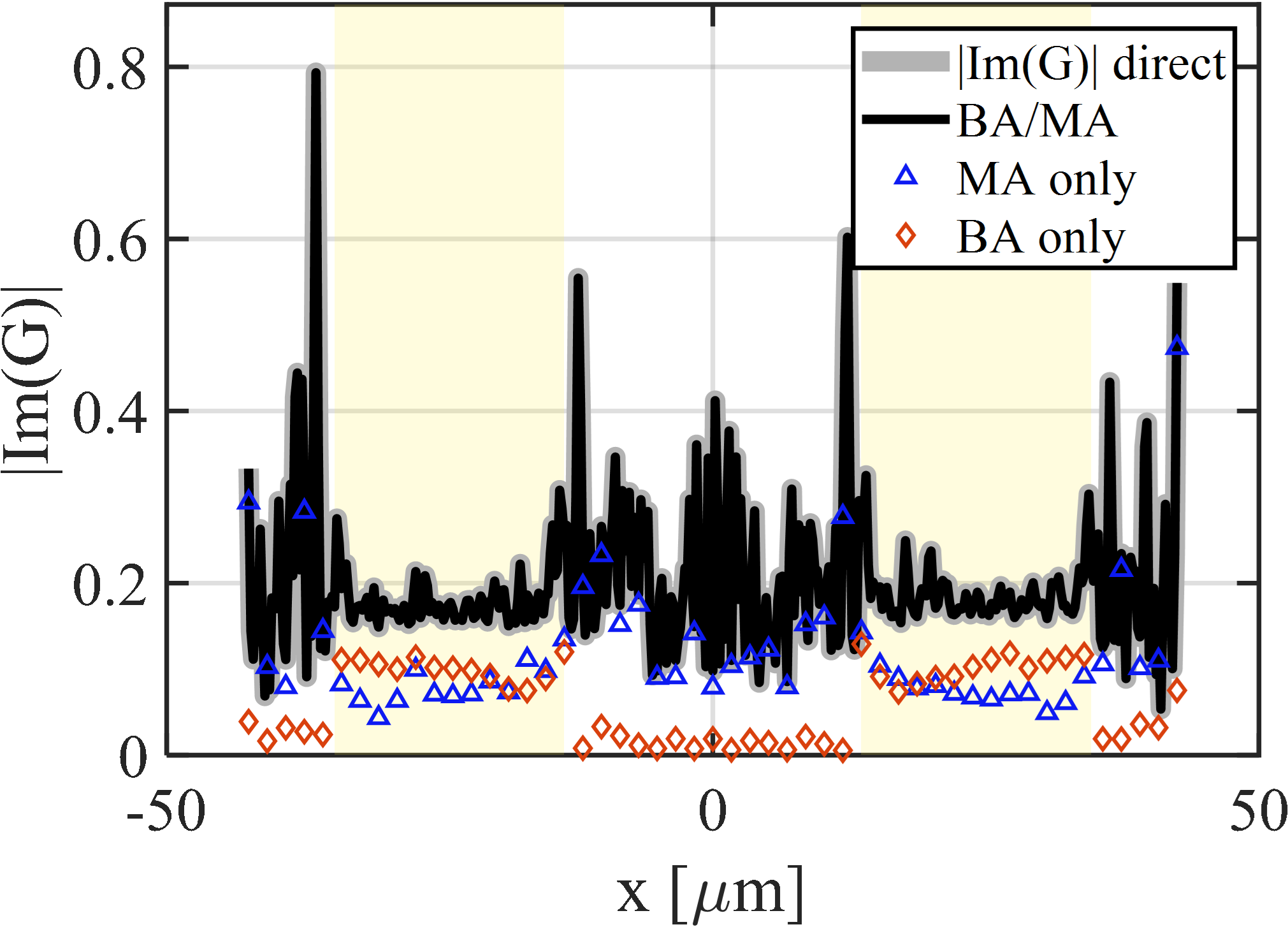}
        \vspace{-1.0ex}

        \textbf{(d)}
    \end{minipage}

    \caption{
    Local Green-function spectral-density reconstruction along a horizontal cut of the PIC devices.
    In all panels the source and observation points coincide,
    \(\mathbf r_s=\mathbf r=(x,y_{\rm cut})\), with \(y_{\rm cut}=0.35\,\mu{\rm m}\), and \(x\) is swept across the full computational domain.
    (a,b) Nominal material-loss parameters for the broadband directional coupler and the ring-loaded interferometric splitter, respectively.
    (c,d) The same two structures with the imaginary part of the permittivity increased by a factor of \(20\).
    Gray curves show the direct point-source evaluation of \(|{\rm Im}\,G(\mathbf r,\mathbf r;\omega_c)|\), and black curves show the full reconstruction from the combined BA/MA channel sum.
    Blue triangles and red diamonds denote the MA-only and BA-only contributions, respectively.
    Shaded regions mark intervals where the horizontal sweep line intersects high-index waveguide cores.
    }
    \label{fig:SM_port_ImG_reconstruction_combined}
\end{figure*}

To verify the completeness of the dissipative-channel representation in the PIC port-boundary simulations, we reconstruct the local Green-function spectral density along a fixed horizontal material cut. 
For each device, the source and observation points are taken to be identical,
\(\mathbf r_s=\mathbf r=(x,y_{\rm cut})\), with \(y_{\rm cut}=0.35\,\mu{\rm m}\), and \(x\) is swept across the full finite-element domain. 
This cut passes through the upper waveguide cores in the coupling sections; the shaded regions in Fig.~\ref{fig:SM_port_ImG_reconstruction_combined} indicate the intervals where the sampled points lie inside high-index material. 
The direct quantity is the local diagonal spectral function \(|{\rm Im}\,G(\mathbf r,\mathbf r;\omega_c)|\), evaluated by point-source radiation problems.

Figure~\ref{fig:SM_port_ImG_reconstruction_combined}(a,b) compares the direct point-source evaluation with the reconstruction obtained by summing the medium-assisted (MA) and boundary-assisted (BA) channel contributions for the nominal material-loss parameters. 
The agreement within numerical precision between the direct result and the combined channel sum confirms that the retained material and port-boundary channels form a complete representation of the lossy port-boundary Maxwell problem. 
The partial MA-only and BA-only reconstructions show how the local fluctuation spectrum is partitioned between material absorption and open-boundary exchange; they are not independent approximations. 
Only the full BA/MA sum reproduces the spectral density.
Figure~\ref{fig:SM_port_ImG_reconstruction_combined}(c,d) shows the same reconstruction after increasing \(\epsilon''_{\rm core}\) and \(\epsilon''_{\rm clad}\) by a factor of \(20\). 
This calculation is included solely to visualize how stronger material absorption redistributes the fluctuation weight between MA and BA channels. 
As expected, the MA contribution becomes more prominent, but the direct Green-function spectral density is still recovered only by the combined BA/MA channel sum. 
Thus changing the relative strength of material absorption changes the channel weights, not the completeness requirement.

\bibliography{sorsamp}